# Constructing the Field of Philanthropic and Nonprofit Studies: Evidence from Citation Networks

**Jin Ai**, PhD, Rutgers University, jin.ai@rutgers.edu.

**Richard S. Steinberg**, PhD, Indiana University, rsteinbe@iu.edu.

**Chao Guo**, PhD, University of Pennsylvania, chaoguo@sp2.upenn.edu.

**Filipi Nascimento Silva**, PhD, Northwestern University, filipi.nascimento@kellogg.northwestern.edu.

## ABSTRACT

Philanthropic and nonprofit studies (PNPS) has grown rapidly as an interdisciplinary field, yet its intellectual structure and boundaries remain only partially visible. This study maps the field by combining journal- and keyword-based retrieval with citation network analysis, natural language processing, and large language models-assisted cluster labeling. Using 60,917 Web of Science articles, we identify major topical communities, examine their structural connections, and assess how well mainstream PNPS journals represent the broader landscape of related scholarship. The results show that PNPS is a loosely connected field dominated by three distinct "power centers": nonprofit organizations, social movements, and voluntary action. The analysis also reveals a broader disciplinary footprint than commonly recognized, extending beyond the social sciences and humanities into areas such as biomedicine and technology. These findings clarify the organization of PNPS and highlight opportunities for stronger integration across research domains.

**Keywords**: Philanthropic and Nonprofit Studies, Citation Networks; Intellectual Structure; Natural Language Processing, Large Language Models

## INTRODUCTION

Over the past century, especially since the 1970s, the distinct research field of Philanthropic and Non-Profit Studies (PNPS) has emerged to understand the organization, actors, and practices of the philanthropic and nonprofit sector (Barman, 2017; Cnaan, 2025; Powell & Steinberg, 2006; Steinberg, 2004). Its growth is evidenced by the proliferation of dissertations, publications, research centers, and educational programs (Bushouse et al., 2023; Cnaan, 2025; Mirabella, 2007; Rooney & Burlingame, 2020; Shier & Handy, 2014; Walk & Andersson, 2020). These efforts come from various disciplinary perspectives, covering a broad spectrum of topics such as regulation of nonprofits (in the disciplines of law and public policy), nonprofit administration (in the disciplines of management, business, and public administration), prosocial behaviors (in the disciplines of social psychology and sociology), collective action (in the disciplines of sociology, political science, and economics), reciprocity and gift giving (in the discipline of anthropology).

Philanthropy is a contested social phenomenon (Breeze, 2021; Daly, 2012), and scholars studying closely related phenomena frequently draw on distinct vocabularies, journals, and citation practices, limiting intellectual cross-fertilization. Disciplinary publication norms shape the framing of research questions and citation practices. The rapid expansion of scientific publishing (Wang & Barabási, 2021) amplifies this fragmentation. Emerging research areas within the field encounter difficulty gaining visibility and integrating with existing work, leaving them at risk of becoming the proverbial "inch-deep and mile-wide."

The field has been mapped using qualitative systematic reviews and quantitative bibliometric analyses that uncover thematic emphases, disciplinary composition, and intellectual structures (LePere-Schloop & Nesbit, 2022; Ma & Konrath, 2018; Shier & Handy, 2014). These studies operationalize the field from lists of core journals or curated lists of keywords. As a result, existing mappings provide complementary but differently bounded representations of the field, highlighting different areas of scholarship as central or peripheral. It is difficult to assess how the intellectual core of PNPS connects to the broader interdisciplinary literature from existing studies, so this study concerns three related questions: (1) What are the major topical communities within PNPS? (2) How are these areas structurally connected through citation relations? (3) To what extent do mainstream PNPS journals represent the broader landscape of scholarship connected to the field?

The current study builds on prior efforts by combining journal- and keyword-based retrieval with citation network analysis. We assemble a broad corpus of publications from the Web of Science (WoS) and construct a citation network, enabling us to identify clusters of research within the field. We use natural language processing (NLP) and a large language model (LLM) to generate concise titles and descriptions. We analyze relationships among clusters and assess the extent to which mainstream PNPS journals represent the broader landscape of PNPS. Our findings show that PNPS is not a coherent field organized around a single intellectual core, but rather a field dominated by three distinct "power centers" that are weakly connected with each other: nonprofit organizations, social movements, and voluntary action. The analysis also reveals a broader disciplinary footprint than commonly recognized, extending beyond the social sciences and humanities into areas such as biomedicine and technology.

Our study advances existing literature in both *substantive* and *methodological* terms. First, by adopting an inclusive mapping approach that combines both journal- and keyword-based retrieval strategies, we significantly expand the scope of the mapping, generating a record-large corpus consisting of over 60,000 articles with half-million references spanning a 124-year period. Second, we model PNPS as a unified citation network that simultaneously includes core outlets and adjacent scholarship. This citation-connected network allows us to examine how major and peripheral topical communities emerge from patterns of citation among individual studies and compare this wider ecosystem of publications with the core discourse within mainstream PNPS journals. Third and methodologically, our analysis follows a multi-stage pipeline that integrates citation network analysis, NLP, and LLM-assisted interpretation to map the intellectual structure of PNPS. As such, our study illustrates how computational bibliometric approaches can enhance the analysis of large and dispersed bodies of scholarship.

In the next section, we review the central concepts commonly used to define the field, then synthesize prior mapping efforts and situate our contribution to this literature. Next, we detail the data, sampling strategy, and analysis pipeline used to construct the citation network and find topical clusters within it. Next, we present findings on publication growth, citation impact, disciplinary composition, the intellectual landscape, and the relationship between mainstream journals and the broader ecosystem. We conclude with implications for field development and directions for future research.

## LITERATURE REVIEW

### Key Concepts and Conceptual Variation

PNPS, as an interdisciplinary research field,[1] is subject to continuous evolution shaped by "prevalent theories, discourses, and practices" (Daly, 2012, p. 548). The field of PNPS emerged from the previously separate fields of nonprofit organizations and voluntary action research. The Program on Non-Profit Organizations (PONPO) at Yale University, founded in 1978, played a central role in the development of the first field. PONPO's mission was "to foster interdisciplinary research aimed at developing an understanding of nonprofit organizations and their role in economic and political life" (PONPO Archives, 2026). The Association of Voluntary Action Scholars (AVAS), founded in 1971, played a central role in developing the second field, defining voluntary action as "all kinds of noncoerced human behavior, collective or individual, that is engaged in because of a commitment to values other than direct, immediate remuneration. Thus, voluntary action includes "a focus on voluntary association, social movements, cause groups, voluntarism, interest groups, pluralism, citizen participation, consumer groups, participatory democracy, volunteering, altruism, helping behavior, philanthropy, social clubs, leisure behavior, political participation, religious sects, etc." (Journal of Voluntary Action Research, 1985). Efforts to bring the two fields together culminated in the renaming of AVAS to the Association for Research on Nonprofit Organizations and Voluntary Action (ARNOVA) and its journal to the *Nonprofit and Voluntary Sector Quarterly (NVSQ)* in 1989.

Despite such efforts at convergence, key concepts in the field are often interpreted differently across disciplinary traditions, reflecting its diversity. Reviewing these concepts helps clarify the range of perspectives that define the field and provides context for understanding why mapping its intellectual structure is analytically challenging.

**Charity and Philanthropy.** Although often used interchangeably in practice, charity and philanthropy carry distinct analytical connotations (Cheek et al., 2015; Katz, 2006). Charity is typically associated with direct assistance and immediate relief, whereas philanthropy often denotes more strategic, institutionalized, and longer-term approaches to social problem-solving across domains such as education, arts, and religion. Charity commonly refers to acts of giving (e.g., donating money, goods, or services) directly toward those in need, while philanthropy emphasizes organized, systematic efforts to produce lasting social change. Philanthropy may include both formal and informal giving, mediated through organizations (nonprofit or not nonprofit) or enacted independently (Anheier & Leat, 2006; Payton & Moody, 2008).

**Nonprofit.** Nonprofit organizations are typically defined by a nondistribution constraint, which restricts the distribution of profits to those in control of the organization (Hansmann, 2016; Steinberg, 2006). “Nonprofit” is an awkward term because it defines part of the field by what it is not (not-for-profit) rather than what it is (Lohmann, 1989). Moreover, many nonprofits rely heavily on commercial revenue and paid labor rather than donated resources, complicating simple associations between nonprofit status and charitable intent. This is particularly notable in cases such as nonprofit hospitals, which may operate on commercial principles and challenge conventional understanding of “charity.” The overlapping use of “nonprofit,” “charity,” and “philanthropy” to refer both to organizations and to behaviors further blurs conceptual boundaries within the field.

**Altruism.** Philanthropy is partly motivated by altruism, but the relationship is complicated by varying definitions across the disciplines. Anthropological perspectives emphasize reciprocity and social obligation, sometimes questioning whether pure altruism exists (e.g., Derrida, 1992). Psychologists (Batson & Powell, 2003) define altruism as behavior intended to benefit others as opposed to “egoism,” behavior intended to benefit the self, a distinction originally developed by philosophers (e.g., MacIntyre, 1967). They distinguish altruism from prosocial behavior, noting that prosocial acts may benefit others without being motivated by altruistic intent. Behavioral economists examine related constructs such as inequality aversion, warm glow, and reciprocity within models of other-regarding preferences (Dimick et al., 2018). Evolutionary biology situates cooperation and self-sacrifice within adaptive processes tied to group survival (Henrich & Muthukrishna, 2021). These differing definitions generate distinct theoretical traditions that inform PNPS research in diverse ways.

**Voluntarism.** In the historical Western colonial context, philanthropy typically referred to monetary donations. Over time, the concept evolved to encompass non-monetary forms of giving. Voluntarism/volunteerism/voluntary action emphasizes the engagement of individuals and groups in addressing social issues and contributing time, effort, and advocacy. While voluntarism and volunteerism are often used interchangeably, volunteerism often emphasizes the workforce aspect of nonprofit organizations. In contrast, voluntarism carries broader connotations of voluntary civic action and individual autonomy. David H. Smith (2013, 2016) proposed the term ‘voluntaristic’ as an alternative umbrella concept for the field, arguing that philanthropy may be “too narrow, financially focused, and elitist in its connotations in the English context” (Smith, 2016, p. 9). This framing broadens the field beyond monetary giving to encompass participation, association, and civic engagement.

**Civil Society.** Civil society is an amorphous term that may refer to a part of society, a type of society, or the public sphere (Edwards, 2009). As a part of society, it denotes associational life. As a type of society, it refers to democratic norms and the obligations of citizenship. The public sphere includes grassroots and other advocacy organizations, voting behavior, and the resolution of disputes through civil means rather than violence. The intersection of civil society and philanthropy emphasizes the political and legal function of philanthropic acts. In some cultures and from the perspective of civil society scholars, nonprofits and giving behaviors are part of the civil society sphere. Overall, while philanthropy centers on giving, civil society focuses on engagement. Civil society provides the context in which philanthropy operates, whereas philanthropic activities strengthen civic space (Lang, 2021; LeRoux & Feeney, 2015).

**Social Economy.** The term social economy is more commonly used in European and Canadian scholarly communities to characterize the PNPS field. This term refers to private entities that engage in economic and social activities in the interests of their members or, more generally, society. Nonprofit organizations are part of the social economy, but it also includes cooperatives, mutual benefit societies, worker-managed firms, and associations that distribute profits based on membership through democratic and participative processes, with an emphasis on the "ethos of solidarity or mutuality" (Anheier & Salamon, 2016, p. 91). This framing expands the field beyond nonprofit legal status to include hybrid and collective forms of organization oriented toward social purpose.

Together, these concepts illustrate the interdisciplinary breadth of the field and the variation in how core phenomena are defined and theorized. Such conceptual diversity enriches the PNPS field but also disperses related scholarship across different disciplinary traditions and publication venues. Mapping the field, therefore, requires attention not only to definitions but also to how these strands are organized and connected within the broader scholarly landscape.

## Knowledge Development and Prior Mapping Efforts

As citation databases expanded, bibliometric and review-based analyses have become an important way to examine how scholarly fields evolve, organize, and consolidate over time (Ball, 2017; Garfield et al., 1978; Radicchi et al., 2017). In PNPS, several such analyses have identified dominant themes, disciplinary influences, institutional growth, and patterns of intellectual development. Prior mapping efforts generally relied on two primary strategies: journal- and keyword-based retrieval. However, these efforts differ in how they operationalize the boundaries of PNPS, which produces partially overlapping portraits of what appears central, peripheral, or external to the field.

**Journal-Based Analysis.** A landmark journal-based analysis is Ma and Konrath (2018), who conducted one of the largest citation studies of PNPS. Using a curated set of 19 PNPS journals drawn from the 61 journal lists developed by David Horton Smith (2013), they applied bibliographic coupling and co-citation techniques to examine thematic clustering and intellectual cohesion across nearly a century of scholarship. Their analysis shows that PNPS knowledge production has grown in quantity, cohesion, and methodological sophistication, while remaining limited in geographic and cultural diversity. They also identified several major clusters (e.g., volunteering theories and behaviors, organizational effectiveness, and accountability) and demonstrated increasing internal cohesion within PNPS over time. These findings show that PNPS contains identifiable topical communities rather than a diffuse set of loosely connected studies, and that internal citation ties among core journals have strengthened as the field matured. However, their cluster identification was conducted primarily across three core journals (i.e., *NVSQ*, previously known as *Journal of Voluntary Action Research*), *Nonprofit Management and Leadership (NML)*, and *VOLUNTAS: International Journal of Voluntary and Nonprofit Organizations*). While this design offers a clear view of internal consolidation, it limits insight into how PNPS connects to adjacent literatures published outside these outlets.

Walk and Andersson (2020) complement this perspective by surveying nonprofit scholars about publication practices and perceived journal centrality. Their findings show that PNPS scholars publish across a much wider range of disciplinary journals than mainstream outlets alone, reinforcing the idea that the intellectual footprint of the field extends beyond what appears in the core journals. While important, an analysis of how those dispersed publications are connected through citation relationships is not included in the scope of this work.

**Keyword-Based Analysis.** Keyword-based approaches retrieve scholarship based on conceptual content rather than publication venue. Shier and Handy (2014), for example, used keyword-based retrieval to analyze trends in nonprofit-focused doctoral theses. Their study documents sustained growth in graduate research and increasing thematic diversification across five main areas (i.e., resources, organizational effectiveness and performance, organizational development, intra-organizational context, and collaboration). While this work provides insight into emerging research trajectories and the training pipeline of the field, its keyword design emphasizes formal nonprofit activities and provides limited insights into the informal aspects of nonprofit practices and the relationships among the identified thematic clusters.

LePere-Schloop and Nesbit (2022) combine journal- and keyword-based retrieval and conduct citation analysis across the resulting dataset. They show that PNPS engages multiple disciplinary domains and that citation flows between the three core journals and external outlets are unbalanced. Their analysis also reveals that many publications using nonprofit-related terminology do not substantially cite core nonprofit journals, suggesting partial fragmentation or parallel development across domains. This design moves beyond single-journal analysis and provides an important view of how the core journals connect to surrounding literatures.

Their corpus is constructed using a broad set of predefined nonprofit-related keywords, including sectoral identifiers (e.g., nonprofit, third sector, civil society), organizational forms (e.g., nongovernmental, voluntary association, social enterprise), and selected altruism concepts (e.g., philanthropy, generosity, charity, altruism, civic engagement), combined with filtering rules on journal scope. As part of this process, certain high-noise terms (e.g., voluntary) are restricted or excluded to reduce false positives. While these decisions improve precision, they may also exclude relevant scholarship. More broadly, this strategy emphasizes formal sectoral labels and selected altruism concepts, which helps capture a wide range of nonprofit-related research but may underpresent areas where nonprofit activity is discussed without explicit sectoral terminology, or where related concepts are embedded in other topical umbrellas, such as funding mechanisms (e.g., grantmaking, donor-advised funds, impact investing), micro-level giving behaviors (e.g., donation, volunteering, reciprocity), institutional and legal forms (e.g., 501(c)(3), foundations, hybrid entities), and cross-cultural or religious giving practices (e.g., zakat, tzedakah, waqf). In addition, the combination with journal-based filtering means that the scope of the dataset is shaped by a series of inclusion and exclusion (e.g., removing articles based on journal categories such as medicine or engineering) decisions made during sample construction.

Their analysis is also organized primarily by journals and disciplinary categories, with particular emphasis on the position and influence of the three flagship PNPS journals. As a result, the structure of the field is interpreted relative to these core outlets, rather than derived from article-level citation relationships. This design provides a clear view of disciplinary contributions and journal positioning, while leaving open questions about how research areas cohere into citation-linked communities when the field is modeled as an integrated network of articles.

Other keyword-based reviews focus more narrowly on specific subfields. For instance, Gazley and Guo (2020) provided a systematic review of nonprofit collaboration research, identifying dominant theories, methods, and research gaps. Such work offers deep insight into particular areas, though it does not aim to map the structural configuration of PNPS as a whole.

Together, these studies show that PNPS is multidisciplinary and that important scholarship appears beyond its core journals. They also reveal uneven integration between mainstream PNPS discourse and adjacent literatures.

**Institutional and Infrastructural Perspectives on Field Formation.** Another stream of work examines PNPS not primarily through citation patterns but through institutional development. Smith (2013) documented the historical growth of research associations, journals, and scholarly communities, tracing the formation of AVAS, ARNOVA, and related infrastructure that support the field. His work demonstrates that PNPS has developed many of the institutional characteristics of a recognized academic discipline. Similarly, historical and reflective accounts (e.g., Bushouse et al., 2023; Cnaan, 2025) emphasize the emergence of PNPS through academic programs, research centers, professional associations, and cross-national networks. These perspectives show how the field has institutionalized and expanded over time beyond journal publications alone. While this line of scholarship established the organizational and professional foundations of PNPS, it does not directly examine how intellectual communities are structured through patterns of scholarly exchange.

**Summary.** As discussed above, existing mapping efforts clarify the growth, thematic concentration, multidisciplinary engagement, and institutional consolidation of PNPS. They demonstrate the presence of identifiable and expanding subfields with established professional infrastructures.

Our study differs from previous work in many ways. Unlike Gazley and Guo (2020), we include the entire field of PNPS and adjacent fields. Unlike Ma and Konrath (2018), we include articles published outside 19 journals. We do not include PhD dissertations like Shier and Handy, but we do include more than 60,000 articles indexed in the WoS. Our paper is most similar to LePere-Schloop and Nesbit (2022): we both use journal and keyword strategies to obtain our sample and employ citation network methods to the data. However, their approach to selecting keywords underrepresents topics such as funding mechanisms, forms of giving, and hybrid organizations. In addition, they eliminate articles published in journals that do not publish many other articles in the field. Hence, four of the five disciplinary domains in the WoS (life sciences and biomedicine, technology, arts and humanities and physical science domains) are significantly underrepresented. Finally, their citation network has journals as nodes, a coarser grid than our article-based nodes. Our broader sample, networked at the article level, produces a unified citation network that includes both the core recognized by those currently working in the field and adjacent scholarship. We construct clusters of core and peripheral topical communities within this network and identify areas of weak integration within the core and between the core and adjacent scholarship.

## DATA AND METHODOLOGY

We used citation data from the WOS-CORE-ESCI 2023 snapshot for its authority, comprehensive coverage, and inclusion of topical classification (Visser et al., 2021). The WoS Core Collection (CORE) and Emerging Sources Citation Index (ESCI) include peer-reviewed journals and established non-peer-reviewed sources across the social and natural sciences, while excluding outlets that do not meet editorial quality standards (Clarivate, n.d.-b). Although its coverage is only one-quarter that of Google Scholar (Martín-Martín et al., 2020), it provides curated, structured metadata. We include only ‘articles,’ including research papers, conference proceedings, and magazine articles, published in academic and trade journals and conferences that have cited references (Clarivate, n.d.-a). The *WoS Journal Citation Reports* categorizes each included journal into at least one of five broad disciplinary domains (Social Sciences, Arts & Humanities, Physical Sciences, Technology, and Life Science & Biomedicine) and at least one of 254 cross-disciplinary subject categories (Birkle et al., 2020).

### Data Sampling

Our sample combines journal- and keyword-based approaches (Figure 1). We began with journal lists compiled by Smith (2013) and Walk and Andersson (2020), eliminated journals not indexed in the WoS, and then each author independently evaluated each journal to determine whether its primary scope aligns with PNPS. If at least two authors agreed, the journal was retained, and our sample begins with all articles appearing in 19 journals (listed in Supplementary Information (SI), section [A]). We then created a keyword list and used it to add articles published outside these journals. To create our list of keywords, we began with the lists used in prior bibliometric studies (Burlingame, 2004; LePere-Schloop & Nesbit, 2022; Shier & Handy, 2014; Smith, 2016). Our focus was on philanthropic actors classified as 501(c)(3) or 501(c)(4) categories (Hall, 2006), excluding other tax-exempt categories. Keywords were eliminated if they added too many irrelevant articles without adding relevant articles not already included from other keywords (for example, "social work," "social welfare," "micro-finance," and "cause-related marketing"). Terms were refined using singular, bi-gram, and paired expressions when necessary (e.g., "gift giving" instead of "giving," or paired searches such as ("giving", "pledge")). Variants consistently associated with false positives were excluded (e.g., "altruistic locking").

**Figure 1. Keywords Curation Process**

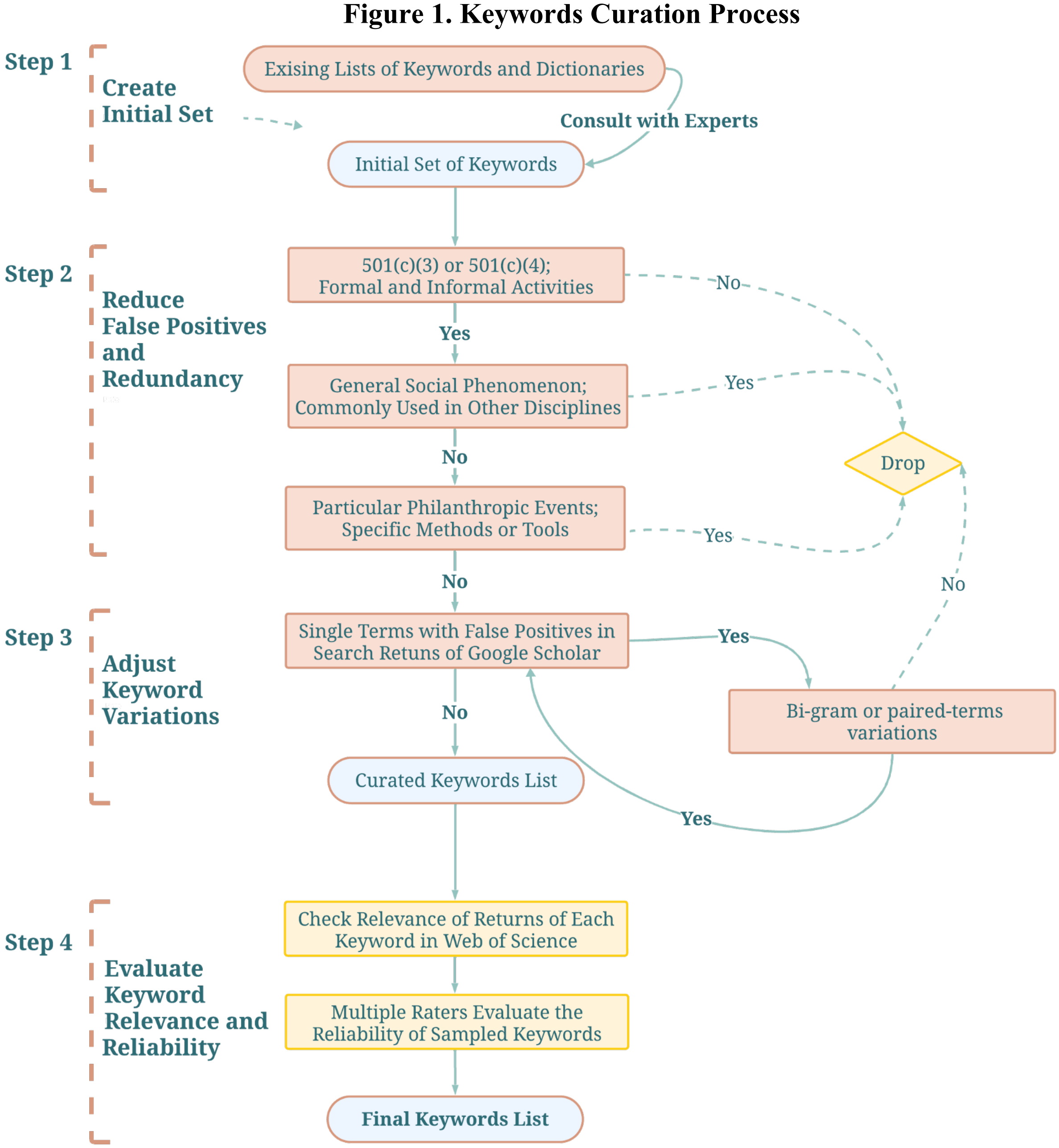

To ensure that the curated keywords effectively captured PNPS scholarship, we conducted two validation procedures. One author reviewed the first 100 returns from these keywords and excluded keywords that retrieved more than about 25% of irrelevant or ambiguous articles. Then, a randomly selected subset of keywords was independently evaluated by three authors, who assessed article relevance based on title, keywords, journal source, and abstract; intercoder reliability was calculated using Gwet's AC1 (Gwet, 2008). The levels of agreement (average AC1 score of 0.93) support the consistency of the keyword list. Our iterative process resulted in a final set of 169 curated keywords. The full list and detailed description of the curation process can be found in SI, Section [B].

**Reliability Check.** Because keyword validation was conducted on a sampled subset rather than the full list, some residual false positives may remain. To assess overall corpus precision, we conducted an additional article-level check. We randomly selected ten articles from each two-decade period (70 per run) and repeated this process three times with shuffled data. We classify articles as relevant or not based on their titles, keywords, journal source, and abstract. The average relevance rate across iterations was 94.3% (Table 1), indicating high overall precision of the combined journal- and keyword-based retrieval. Although some noise may persist, citation-network constructions retain only structurally connected publications, and cluster detections rely on aggregate patterns rather than individual term inclusion. Consequently, isolated false positives are unlikely to meaningfully affect the network structure.

**Table 1. Reliability Checks for the Analytical Sample**

| Iteration | Article Counts | Relevance Rate (%) |
|---|---|---|
| 1st | 70 | 94.29% |
| 2nd | 70 | 90% |
| 3rd | 70 | 98.57% |
| **Average** | **70** | **94.29%** |

## Analysis Strategy

Our analysis follows a multi-stage pipeline that integrates citation network analysis, NLP, and LLM-assisted interpretation to map the intellectual structure of PNPS (Figure 2). First, we construct the citation network and identify topical clusters. Second, we characterize the thematic content of each cluster. Third, we generate descriptive labels and summaries for each cluster. using an LLM. Then we perform additional structural and temporal analyses.

**Figure 2. Analytical Pipeline for Mapping the Intellectual Structure of PNPS**

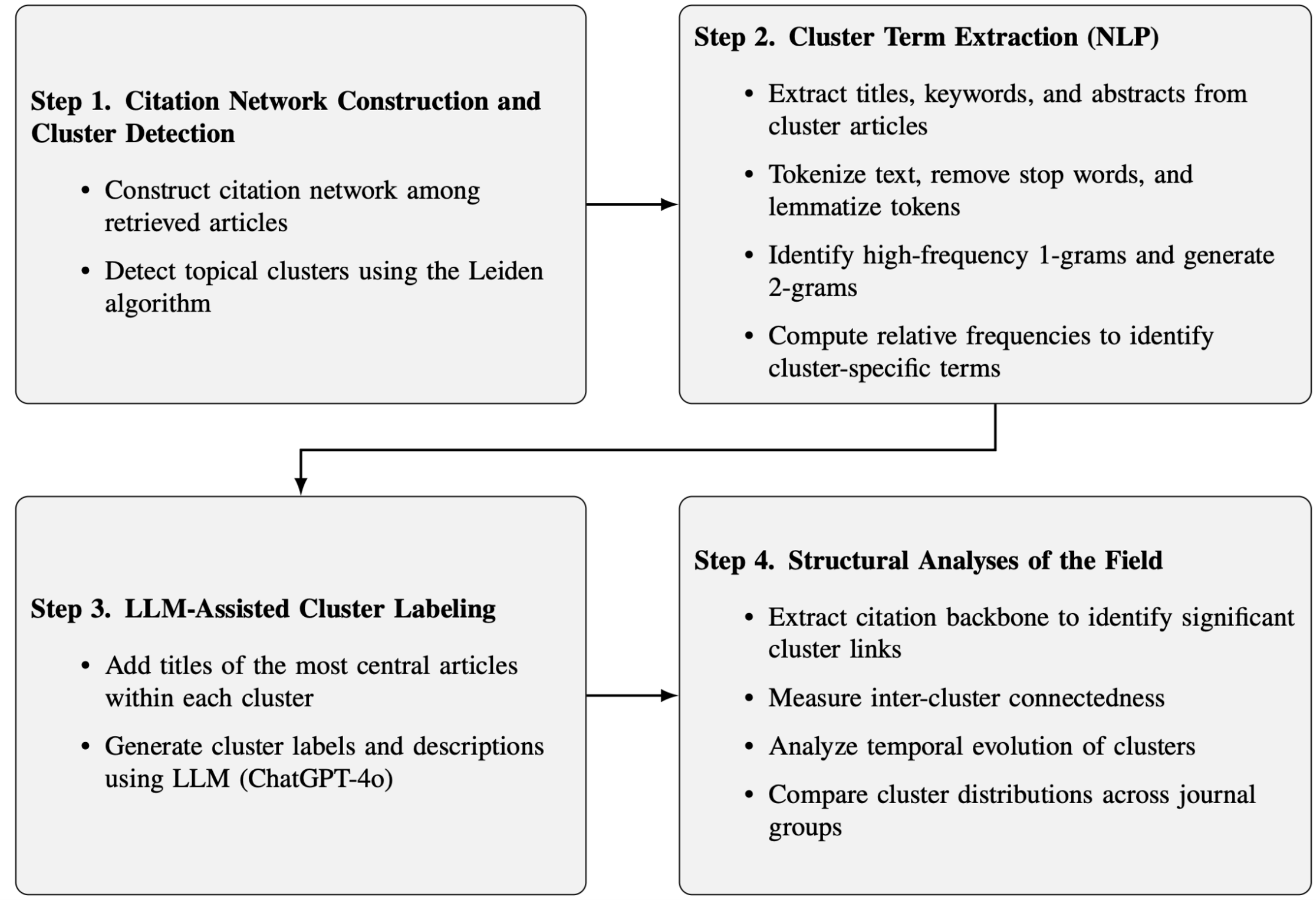

**Citation Network Construction and Cluster Detection.** We first construct the citation network, where nodes represent articles and edges represent citation relationships. Citation networks capture intellectual relationships among research topics by reflecting patterns of knowledge across publications (Borgatti et al., 2009) and displaying interactions of interest-linked groups within a field (Small, 1999). A topical domain, defined as a subfield shared by a group of scholarly works, can therefore be observed through clusters of densely interconnected articles (Fortunato, 2010). To detect these clusters, we apply the Leiden algorithm,[2] which improves upon earlier modularity-based methods by using computationally efficient procedures on large networks to ensure that communities are well-connected (Traag et al., 2019). The resulting clusters represent topical communities that share dense citation relationships. We visualize the non-weighted and non-directional network and its clusters using Helios-Web (Silva, 2023).

**Extraction of Cluster-Specific Terms.** After identifying clusters, we characterize the thematic content of each cluster using an NLP approach adapted from Silva et al. (2016)that integrates textual analysis with citation-based clustering (Brito et al., 2023, 2024). This approach extracts terms that occur disproportionately within a cluster, tokenizes texts by removing stop words, then lemmatizes tokens to generate candidate terms. Initially, both 1-gram and 2-gram tokens were extracted. To reduce noise from long-tail tokens and improve computational efficiency, we retain only the top quartile of 1-gram tokens. Two-gram tokens are retained only when they contain these high-ranked 1-gram tokens. For each token, we compute its relative frequency within and outside the cluster:

$$\textit{Relative Frequency (token i) = In-Cluster Frequency (i) - Out-Cluster Frequency(i)}$$

$$\textit{In-Cluster Frequency (i)} = \frac{Occurrences\ of\ Token\ i\ in\ In{-}Cluster\ Documents}{Total\ Tokens\ in\ In{-}Cluster\ Documents}$$

$$\textit{Out-Cluster Frequency (i)} = \frac{Occurrences\ of\ Token\ i\ in\ Out{-}Cluster\ Documents}{Total\ Tokens\ in\ Out{-}Cluster\ Documents}$$

To further reduce the influence of generic tokens appearing across multiple clusters, we double the weight assigned to the out-cluster frequency. The resulting ranked tokens represent cluster-specific terms. In addition to extracted tokens, we included the titles of the most central articles within each cluster, defined as those with the highest citation connectivity within the cluster, as part of the cluster descriptions. These titles were incorporated without tokenization to preserve the original structure of the language used in the PNPS literature. The top 100 extracted terms and central articles for each cluster are reported in the SI, Section [C].

**LLM-Assisted Cluster Labeling.** After extracting cluster-specific terms, we used an LLM to generate descriptive titles and summaries for each cluster. Specifically, we used ChatGPT-4o accessed via the OpenAI API with default parameters. The model inputs extracted cluster-specific terms and central article titles and produces a concise label and description that represents the cluster's thematic focus. Because LLMs may generate slightly different outputs across iterations, we ran the model five times and manually synthesized the resulting titles and descriptions to produce the final cluster labels. Two rounds of raw outputs are reported in SI, Section [C]. The prompts used and representative outputs are provided in SI, Section [D].

**Structural Analysis of Cluster Relationships.** Following cluster identification and labeling, we conduct additional analyses to examine the structural organization and evolution of the field. First, we extract the backbone of the cluster-level citation using the disparity filter approach of Serrano et al., (2009). The backbone is a simplified network structure that preserves the most important structural features, those unlikely to arise from noise at multiple scales.[3] The backbone illustrates the core structure of the field, including central clusters, connected areas, and peripheral domains.

Second, we measure inter-cluster connectedness using normalized mutual citation links, defined as the number of connections between two clusters divided by the total number of connections within the two clusters. While the backbone highlights the most structurally significant ties, the connectedness measure provides a systematic comparison of connection strength across all cluster pairs, capturing both strong and moderate relationships that may not appear in the backbone. This enables a more complete assessment of how research areas relate to one another across the field.

Third, we analyze the temporal evolution of clusters by tracking their relative publication shares, revealing how different research areas have emerged, expanded, or declined.. Finally, we compare the cluster distributions of articles published in core PNPS journals and a broader set of PNPS-oriented journals with those observed in the full corpus to assess how these publication venues capture the broader intellectual landscape.

## RESULTS

First, we describe the publication growth and citation impact of the retrieved corpus. Second, we examine the domain composition of the literature to assess the multidisciplinary footprint of the field. Third, we map the intellectual landscape by identifying major topical communities and their relationships. Finally, we examine the temporal development of these research areas and compare the topical distributions of mainstream PNPS journals and the full corpus to assess how the core journals represent the field.

### Growth in Publication and Scholarly Impact

A total of 60,917 articles with 520,963 references are identified in the WoS from 1899 to 2023. Figure 3 shows that PNPS-related publications remained sparse until the mid-twentieth century, followed by sustained growth after 1960. This pattern is consistent with the broader expansion of scientific publishing documented across fields (Wang & Barabási, 2021, p. 162). Citation patterns indicate rising aggregate visibility alongside slower growth in per-paper citations. Approximately 21% (12,873) of articles had not received any WoS-recorded citations through 2023. Figure 4 shows that total annual citations increase sharply over time, whereas average citations per publication grow more modestly. Because citation levels vary widely across disciplines and publication periods, these trends are interpreted as descriptive indicators of the changing visibility of PNPS rather than as benchmarks for cross-field impact comparisons.

**Figure 3. Growth of Publications in PNPS, 1899–2023**

**Note:** The blue curve represents the cumulative number of publications over time. The orange curve is the number of articles published each year.

### Domain Compositions

Prior studies characterize PNPS as grounded primarily in social science and humanities disciplines such as economics, management, social work, psychology, political science, public administration, anthropology, history, and religion (LePere-Schloop & Nesbit, 2022; Ma & Konrath, 2018). Figure 5 shows that our retrieved corpus displays a similar pattern, with Social Sciences representing the dominant domain. However, the corpus also contains substantial representation from Life Sciences & Biomedicine, as well as smaller but notable shares from Technology and Physical Sciences domains. This broader disciplinary presence has received little attention in prior mapping efforts, which largely rely on journal-based samples or keyword sets oriented toward nonprofit sector scholarship and therefore primarily captured social science venues. As the cluster-level analysis below (Table 2) illustrates, some topical communities are predominantly social-scientific, whereas others are anchored in biomedical or biologically oriented research traditions. These results suggest that scholarship relevant to PNPS concepts extends into several neighboring research domains that have been less visible in previous bibliometric mappings of the field.

**Figure 4: Trends in Total Citations and Citations per Paper in PNPS**

**Note:** The blue curve represents total annual citations, with an exponential growth rate of 0.158. The orange curve represents average citations per publication, with an exponential growth rate of 0.044.

**Figure 5. Disciplinary Compositions of the PNPS Literature**

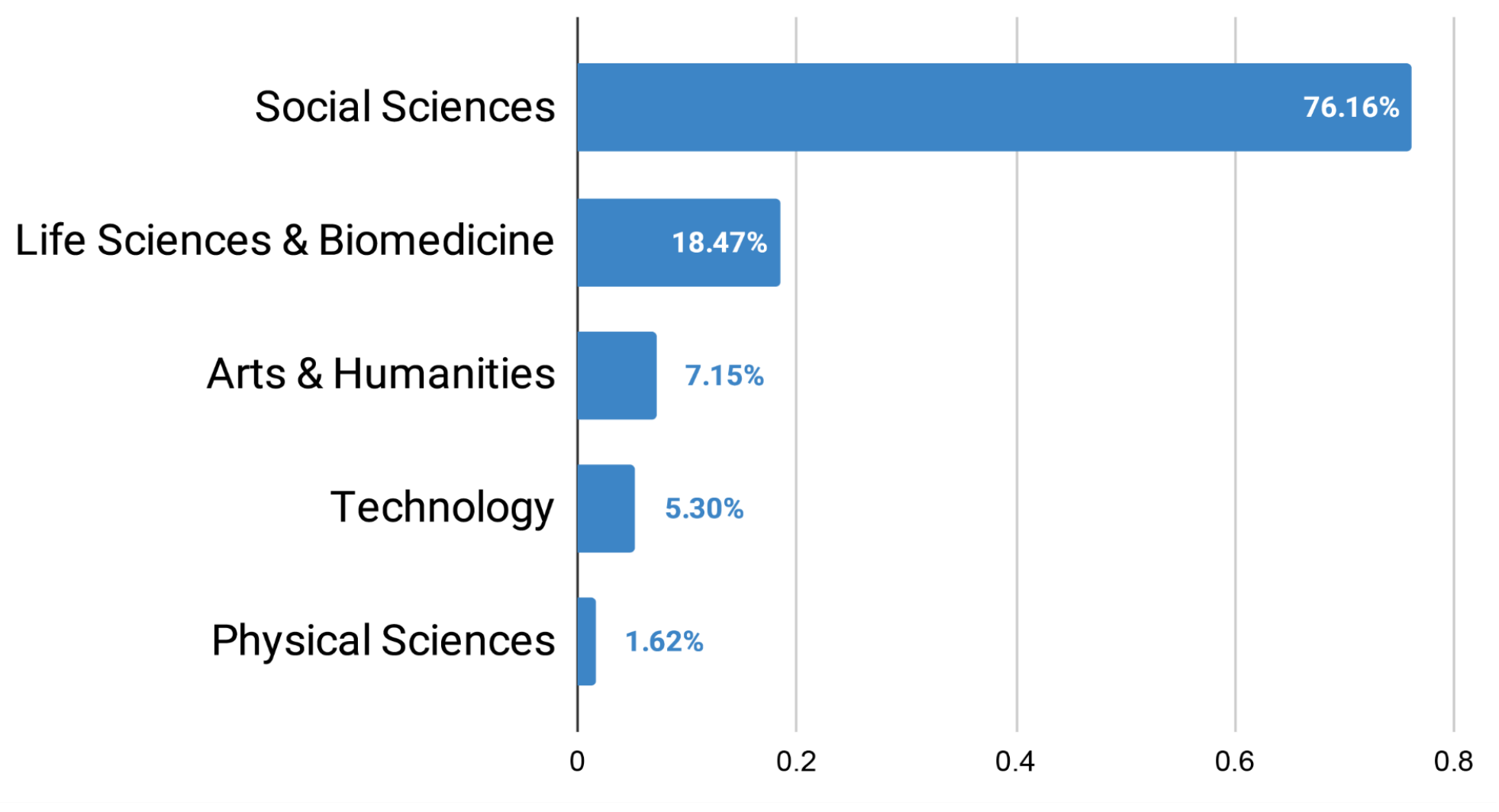

**Note:** Articles may belong to multiple fields, thus total percentage exceeding 100%.

## Intellectual Landscape

Drawing upon the citation relationship, 43,175 out of 60,917 articles either cite or are cited by other articles in this sample, forming the connected component of the citation network (about 30% of articles were isolated and thus played no role in the structural analysis). Using the Leiden algorithm, we identify 20 clusters representing distinct topical communities. Figure 6 presents the intellectual map[4,5] of the field, and Table 2 reports the size and domain composition of each cluster. A description of each cluster follows.

**Figure 6. Intellectual Map of Philanthropic and Nonprofit Studies**

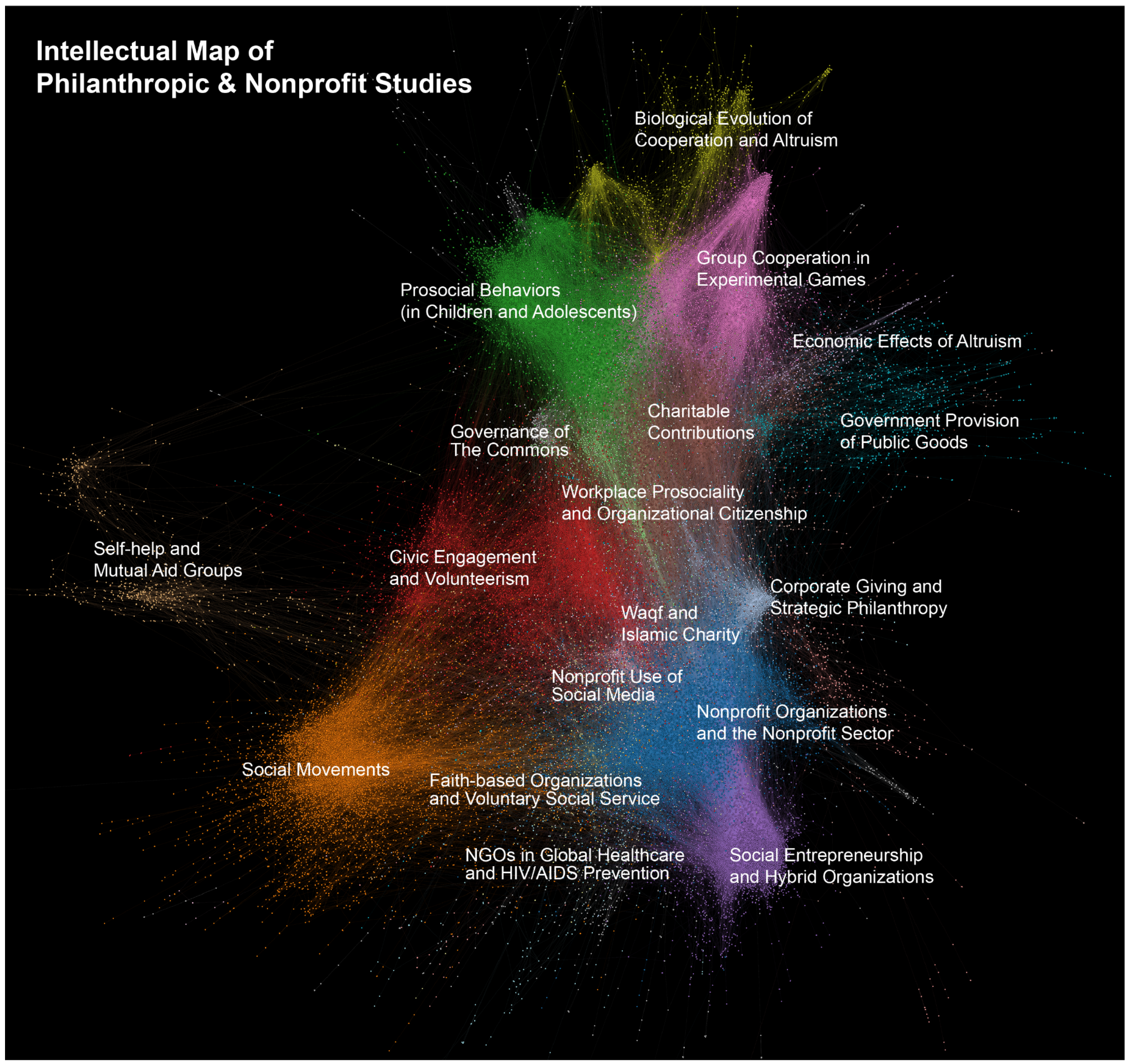

**Table 2. Size and Domain Composition by Cluster**

| | Cluster | Size | Social Sciences | Arts & Humanities | Life Sci. & BioMed | Technology | Physical Sciences |
|---|---|---|---|---|---|---|---|
| 1 | Nonprofit Organizations and the Nonprofit Sector | 6,779 | 91.18% | 1.43% | 9.44% | 2.92% | 0.41% |
| 2 | Social Movements | 6,098 | 90.31% | 4.41% | 12.71% | 2.80% | 0.48% |
| 3 | Prosocial Behaviors (in Children and Adolescents) | 5,290 | 84.35% | 2.34% | 20.09% | 5.92% | 0.30% |
| 4 | Civic Engagement and Volunteerism | 4,039 | 88.56% | 2.67% | 13.44% | 3.99% | 0.50% |
| 5 | Social Entrepreneurship and Hybrid Organizations | 3,890 | 88.41% | 1.21% | 12.62% | 8.25% | 0.28% |
| 6 | Charitable Contributions | 3,465 | 91.17% | 1.96% | 11.37% | 5.86% | 1.21% |
| 7 | Group Cooperation in Experimental Games | 3,239 | 66.56% | 1.45% | 20.90% | 12.63% | 13.00% |
| 8 | Biological Evolution of Cooperation and Altruism | 1,392 | 17.96% | 7.47% | 69.40% | 12.14% | 2.73% |
| 9 | Government Provision of Public Goods | 1,176 | 94.64% | 1.45% | 10.97% | 2.81% | 4.25% |
| 10 | Corporate Giving and Strategic Philanthropy | 946 | 90.06% | 1.90% | 10.78% | 6.24% | 0.63% |
| 11 | Self-help and Mutual Aid Groups | 807 | 57.74% | 1.24% | 57.74% | 2.73% | 0.25% |
| 12 | Workplace Prosociality and Organizational Citizenship | 803 | 86.92% | 1.25% | 13.20% | 9.84% | 0.37% |
| 13 | Waqf and Islamic Charity | 747 | 56.76% | 23.83% | 10.31% | 1.47% | 0.94% |
| 14 | Economic Effects of Altruism | 650 | 87.23% | 1.23% | 16.92% | 8.00% | 8.15% |
| 15 | Governance of The Commons | 600 | 69.83% | 3.33% | 39.83% | 6.50% | 1.67% |
| 16 | Nonprofit Use of Social Media | 528 | 88.26% | 1.33% | 6.44% | 10.23% | 0.76% |
| 17 | Faith-Based Organizations and Voluntary Social Service | 499 | 75.95% | 10.82% | 20.04% | 3.61% | 0.00% |
| 18 | NGOs in Global Healthcare and HIV/AIDS Prevention | 459 | 43.79% | 0.87% | 73.42% | 3.27% | 0.00% |
| 19 | Gift-Giving | 401 | 81.30% | 5.99% | 13.47% | 7.23% | 0.50% |
| 20 | Biological Donations | 299 | 52.17% | 13.71% | 67.22% | 1.34% | 0.33% |

***Cluster 1: Nonprofit Organizations and the Nonprofit Sector***

This cluster examines nonprofit organizations and the nonprofit sector as a whole. At the organizational level, it includes structures, governance, interactions, and practices related to financial sustainability and service delivery within nonprofits. At the sector level, it includes government and non-government partnerships, accountability, and evolution.

***Cluster 2: Social Movements***

This cluster examines social movements, domestic and cross-national, focusing on their emergence, evolution, and impact. Key topics include the strategies and factors shaping mobilization, resistance, and public discourse, such as collective identities, emotions, political opportunities, and traditional and digital media. It also examines the political and cultural contexts in which activism occurs, with particular attention to the interaction between grassroots organizing and broader sociopolitical dynamics and democratic processes.

***Cluster 3: Prosocial Behaviors (in Children and Adolescents)***

This cluster examines prosocial behaviors, primarily in children and adolescents. It includes the measurement of prosociality (using self reports, teacher and parent reports, and longitudinal data), factors affecting the developmental trajectory of prosocial attitudes and behaviors from early childhood through adolescence, influencing factors such as age, gender, and development stage, the influence of the social networks of parents, peers, and teachers, and the interplay between prosocial and antisocial behaviors.

***Cluster 4: Civic Engagement and Volunteerism***

This cluster examines motivations and other determinants of civic engagement and volunteerism (such as age, youth involvement, education, and the role of organizations), the effects of volunteering on the volunteer and society (including effects on social capital, community development, and political participation), and volunteer management (including recruitment, retention, and volunteer satisfaction). It also highlights the role of both formal and informal volunteering in civic engagement across different cultural and national contexts.

***Cluster 5: Social Entrepreneurship and Hybrid Organizations***

This cluster examines social entrepreneurship and hybrid organizations that incorporate both commercial and social objectives. It explores their motivations, ethical challenges, and organizational models, as well as outcomes such as sustainability, innovation, and value creation within the social economy.

***Cluster 6: Charitable Contributions***

This cluster examines motivations for giving (such as warm glow, impure altruism, monetary incentives, image motivation, and tax deductions) and the effectiveness of various fundraising strategies (e.g., targeted appeals, crowdfunding, and marketing). It also includes the socio-economic impacts of charitable giving on public goods and income distribution. Experiments and economic modeling are commonly used to study topics in this cluster.

***Cluster 7: Group Cooperation Experimental Games***

This cluster examines group cooperative behaviors in social dilemma settings from a mostly social science perspective. Experiments (including those based on the public good, dictator, trust, and ultimatum games) are used to test theories of the determinants of cooperation, defection, punishment, and free-riding and to distinguish the motives (including behavioral aspects such as free riding, individual traits, reputation, and social norms and preferences) underlying such decisions.

***Cluster 8: Biological Evolution of Cooperation and Altruism***

This cluster examines evolutionary mechanisms of cooperation and altruism among individuals of various species with a focus on evolutionary game theory, kin selection, cooperative breeding, and group selection. Evolutionary models, simulations, and real-world observations are commonly used. The impacts of population structure, genetic factors, and individual fitness on the evolution of cooperation are also examined.

***Cluster 9: Government Provision of Public Goods***

This cluster examines economic and political factors, such as public spending, local government structure, taxation, voter preferences, and fiscal policies, that shape the cost and optimal government provision of public goods. It includes interactions between private provision by nonprofit organizations and public provision, and the social welfare consequences of public/private provision.

***Cluster 10: Corporate Giving and Strategic Philanthropy***

This cluster examines the determinants (e.g., political connections), strategic alignment with business goals, and benefits of corporate giving. Aspects of benefits include reputation, financial performance, consumer behavior, and stakeholder relationships. Chinese firms receive particular examination.

***Cluster 11: Self-help and Mutual Aid Groups***

This cluster examines self-help and mutual aid groups that assist in substance recovery, well-being, and empowerment, particularly through peer support and group dynamics, across different cultural and regional contexts (e.g., South Asia and Africa).

***Cluster 12: Workplace Prosociality and Organizational Citizenship***

This cluster examines employee motivations (e.g., intrinsic and extrinsic), workplace characteristics (e.g., supervisor trust, leadership styles), and the impacts on job performance of altruistic behaviors at and beyond the workplace. Social exchange theory and structural equation modeling are the primary theories and methods used in this topic.

***Cluster 13: Waqf and Islamic Charity***

This cluster examines mostly Islamic charitable practices with a focus on Waqf, a religious endowment designated for charitable purposes. It highlights the historical and geographical contexts (from the Ottoman Empire to modern applications in countries like Malaysia and Indonesia), institutionalization, development, management, and socio-economic impacts of Waqf in Muslim communities across different centuries and countries.

***Cluster 14: Economic Effects of Altruism***

This cluster examines the economic implications of altruism, particularly intergenerational contexts (e.g., bequests, parental support, and family transfers). It includes the effects of altruism on equilibrium wealth distributions, household utility, and consumption, and on fiscal and welfare policies (e.g., taxation, social security, and economic growth). Economic frameworks frequently used include overlapping generation models and dynastic models.

***Cluster 15: Governance of The Commons***

This cluster examines the tragedy of the commons, as well as collective action and institutional efforts to manage and sustain common resources like forests, fisheries, water, and agricultural land. Elinor Ostrom and Garrett Hardin are influential scholars. The importance of governing the commons is especially pronounced in domains such as agriculture, climate change, and biodiversity.

***Cluster 16: Nonprofit Use of Social Media***

This cluster examines nonprofit social-media use for communication and stakeholder engagement across topics like fundraising, advocacy, cause promotion, and public relations.

Online content strategies, social marketing techniques, and the effectiveness of different platforms (e.g., Twitter and Facebook) are also included.

***Cluster 17: Faith-Based Organizations and Voluntary Social Services***

This cluster examines the religious motives, community engagement, and secular partnerships of faith-based organizations in delivering voluntary social services, especially in areas such as food insecurity, homelessness, and emergency needs in the context of welfare restructuring, neoliberalism, and austerity across different countries (e.g., Canada, the UK, and Australia).

***Cluster 18: NGOs in Global Healthcare and HIV/AIDS Prevention***

This cluster examines the implementation, health outcomes, effectiveness, challenges (e.g., capacity building), and strategies of HIV/AIDS prevention and other healthcare programs delivered by non-governmental organizations and community-based organizations across various regions, particularly Africa and other resource-limited regions. Studies also highlight the important role of capacity building, qualitative assessments, and governmental partnerships in the success of these programs.

***Cluster 19: Gift-Giving***

This cluster examines gift-giving as a social, cultural, and economic practice from a mostly anthropological perspective to understand its tradition, rituals, and reciprocity across different societies (e.g., in China and the Netherlands), including topics regarding social (e.g., interpersonal bonds) and economic (e.g., consumer trust and marketing strategies) impacts, symbolic significance, as well as ethical considerations (e.g., bribery and corruption) of gift giving in contexts of social control, guanxi, bribery, and corruption.

***Cluster 20: Biological Donations***

This cluster examines demographic, social, and psychological determinants that shape organ and gamete donations. It also includes ethical considerations and strategies for increasing donation rates.

**Domain Distribution Across Clusters.** Cluster-level domain composition further illustrates the disciplinary diversity of PNPS (Table 2). While Social Sciences dominate most clusters, several research areas draw heavily from other domains. Life Sciences and Biomedicine account for a large share of publications in Clusters such as 8, 18, and 20, while Technology-related research appears prominently in Clusters 7, 8, and 16. Contributions from the Arts and Humanities are concentrated primarily in Clusters 3 and 20. Physical Sciences are also visible in Cluster 7, reflecting the influence of approaches such as network science and experimental cooperation research. These patterns indicate that while PNPS is anchored in the Social Sciences, several topical communities extend into adjacent disciplinary domains.

## Structural Organization and Intellectual Dynamics of the Field

While the cluster descriptions above outline the major topical areas of PNPS, the citation network reveals how these areas are connected. The citation network results below examine cluster relations, the temporal development of research areas, and the representation of these topics across publication venues, while also highlighting areas that remain weakly integrated.

### *Core Structure of the Field*

The backbone representation (Figure 7) reviews a *hub-connector-periphery* configuration that clarifies how the field is structurally organized. Tie thickness is proportional to the citation weight between clusters after backbone filtering, so thicker ties indicate stronger retained inter-cluster connections. The number of articles in a cluster determines node size..

**Figure 7. Backbone Structure of PNPS Citation Network at the Cluster Level**

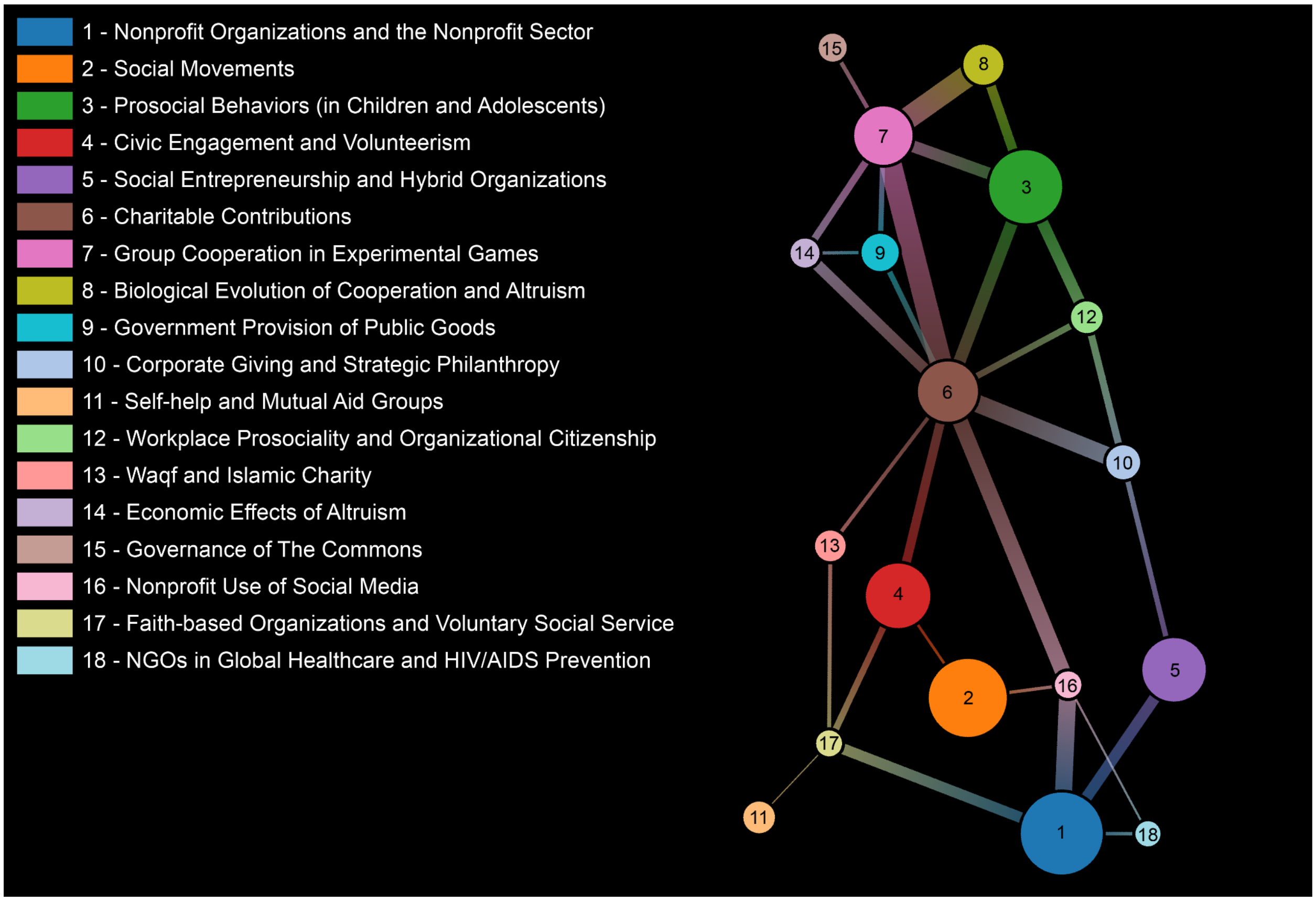

At the center of the network, Cluster 6 (charitable contributions) and Cluster 7 (group cooperation in experimental games) jointly function as *hubs*, but in distinct ways. Cluster 6 maintains the largest number of significant citation ties and serves as the primary coordinating node, directly connecting to multiple major clusters, including 3, 4, 7, 9, 10, 12, 13, 14, and 16. Cluster 7, while slightly less central in degree, occupies a strategic bridging position, linking Cluster 6 to Clusters such as 3, 8, 14, and 15. Surrounding the hubs are a set of clusters that form the main pathways through which the central structure extends outward. These clusters function as *connectors*, linking the hubs to more specialized or distant regions of the network. In particular, Clusters 3 (prosocial behavior in children and adolescents), 16 (nonprofit use of social media), and 17 (faith-based organizations and voluntary social service) occupy intermediate positions and connect clusters that would otherwise be more weakly integrated. These clusters facilitate cross-domain linkages and connect different levels of analysis within the field.

At the *periphery*, Clusters 1, 2, 5, 8, 11, 13, 15, and 18 exhibit relatively few significant ties and connect to the broader network primarily through one or two intermediary nodes. These clusters represent areas that remain more weakly connected with the hubs.

Now, when node size is taken into consideration, Clusters 1 and 2 stand out from the periphery group: they are the two largest clusters in the network, but they are not directly connected with the hubs. Cluster 1, which conventionally represents the core domain of nonprofit organizations and sectoral governance, is larger than the two hubs combined. Cluster 2, which consists of social movement literature, is nearly the size of the two hubs combined. As such, it would be problematic to consider such clusters peripheral in a substantive sense, even if they appear to be peripheral in the network. Also notable are Cluster 11 (self-help and mutual aid groups) and Cluster 15 (governance of the commons): not only are they among the smallest in size, but they are also connected with the broader PNPS network through only one link.

Considering both inter-cluster ties and node strength, the backbone suggests that PNPS is not a coherent field organized around a single intellectual core. Rather, it is a world of “three kingdoms”: a field that is dominated by three distinct spheres of influence. The first two “kingdoms,” Clusters 1 and 2, have each accumulated an enormous body of literature, with strong ties within the cluster but much weaker connections with the rest of the network. The third “kingdom,” broadly labeled as voluntary action, is essentially an alliance of small and closely linked clusters, led by the dual hubs, Clusters 6 and 7, and strongly backed up by Cluster 3. Strong ties from Cluster 6 to Clusters such as 4, 9, 10, 12, and 14 indicate that giving-related processes link domains including civic engagement, public goods provision, corporate philanthropy, workplace prosociality, and economic models of altruism. Cluster 7, in turn, connects these domains to experimental, evolutionary, and theoretical transitions, extending PNPS into broader social science and interdisciplinary research on cooperation.

### *Interconnectedness Across Clusters*

Figure 8 presents the heatmap of normalized citation connections among clusters, illustrating the relative intensity of intellectual exchange across areas. Several cluster pairs display noticeably stronger citation exchange than the rest of the network. The most prominent concentrations appear among Clusters 6 (charitable contributions), 7 (group cooperation in experimental games), and 8 (biological evolution of cooperation and altruism),where mutual citation links are stronger than most other cluster pairs This pattern indicates that scholarship on charitable contributions, experimental cooperation, and evolutionary perspectives on altruism draws on shared theoretical frameworks and empirical findings.

**Figure 8. Heatmap of Normalized Citation Interconnectedness Across PNPS Clusters**

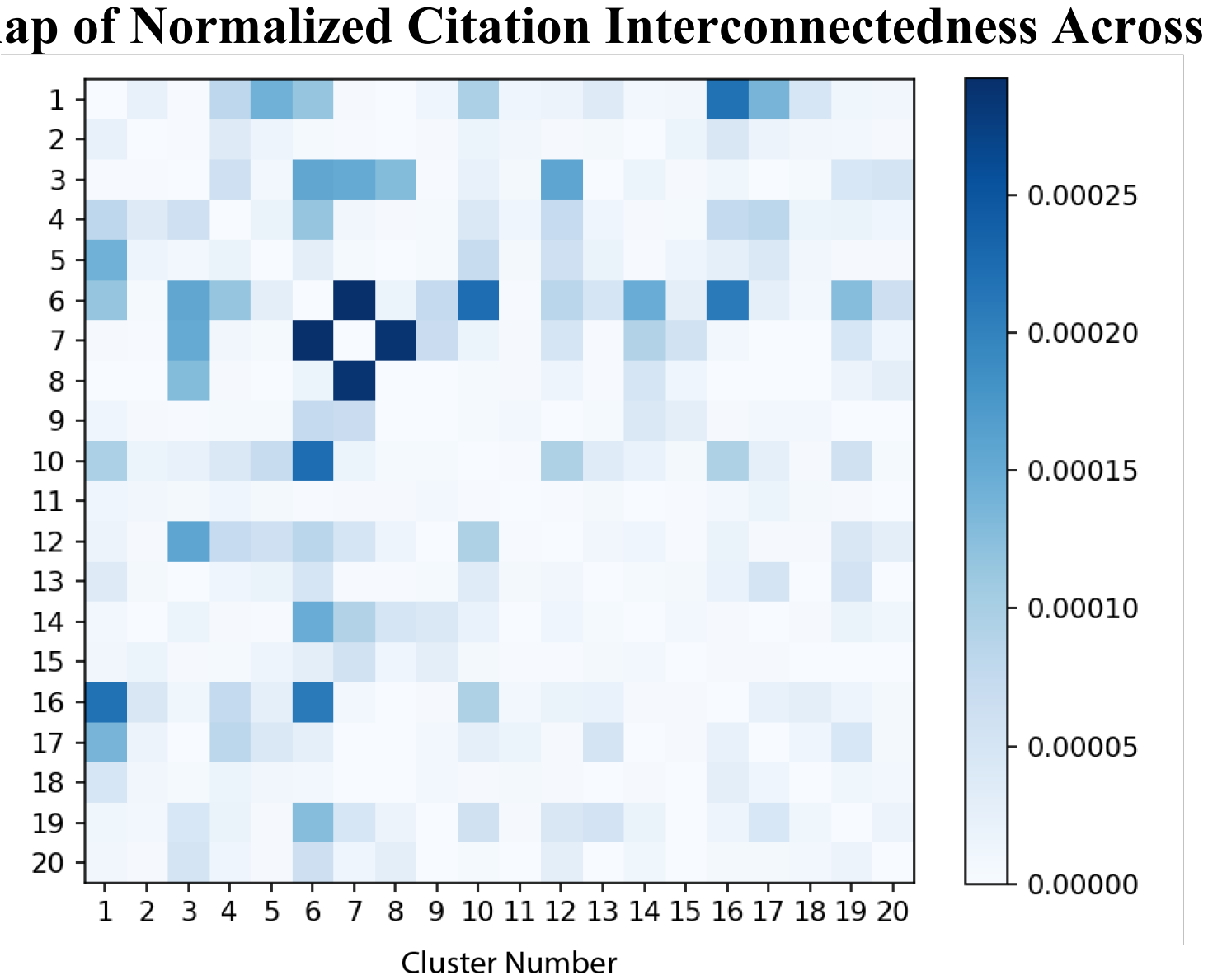

Strong connections are also visible between Clusters 6 and 10, and between Clusters 6 and 16, suggesting that research on charitable giving interacts closely with scholarship on corporate philanthropy and nonprofit communication. Similarly, strong connections are observed between Cluster 1 and Cluster 16. These relationships highlight the key role of these PNPS topics in connecting multiple strands of research within the field. Outside the central region, most cluster pairs exhibit much weaker citation exchange, appearing as lighter cells in the matrix. These connections indicate that many research areas interact with the broader PNPS literature intermittently and remain less integrated into the central scholarly conversation. Overall, the heatmap complements the backbone analysis by showing that the hub-centered structure of the field is accompanied by uneven levels of interaction among research areas, with the most intensive exchange occurring within a narrow core of clusters.

***Temporal Evolution of Research Areas***

Figure 9 traces the temporal evolution of the clusters, showing how their relative publication shares have changed. Several clusters (i.e., Clusters 1, 2, 3, and 4) emerge early and maintain substantial presence throughout the field development, suggesting that these areas form the historical foundation of PNPS scholarship. Other clusters expand more noticeably in later decades. Clusters 5 and 7 show sustained growth beginning in the 1990s and continuing into the 2000s and 2010s, reflecting the increasing prominence of research on social entrepreneurship and experimental approaches to cooperation. Clusters 16 and 17 are more recent, indicating newer research directions that extend PNPS into adjacent areas, such as nonprofit and organizational practices. In contrast, several clusters (e.g., Clusters 8, 9, and 10) were more prominent in earlier decades but now represent smaller portions of the literature. These trajectories suggest shifts in scholarly attention as new research agendas emerge.

**Figure 9. Evolution of Clusters**

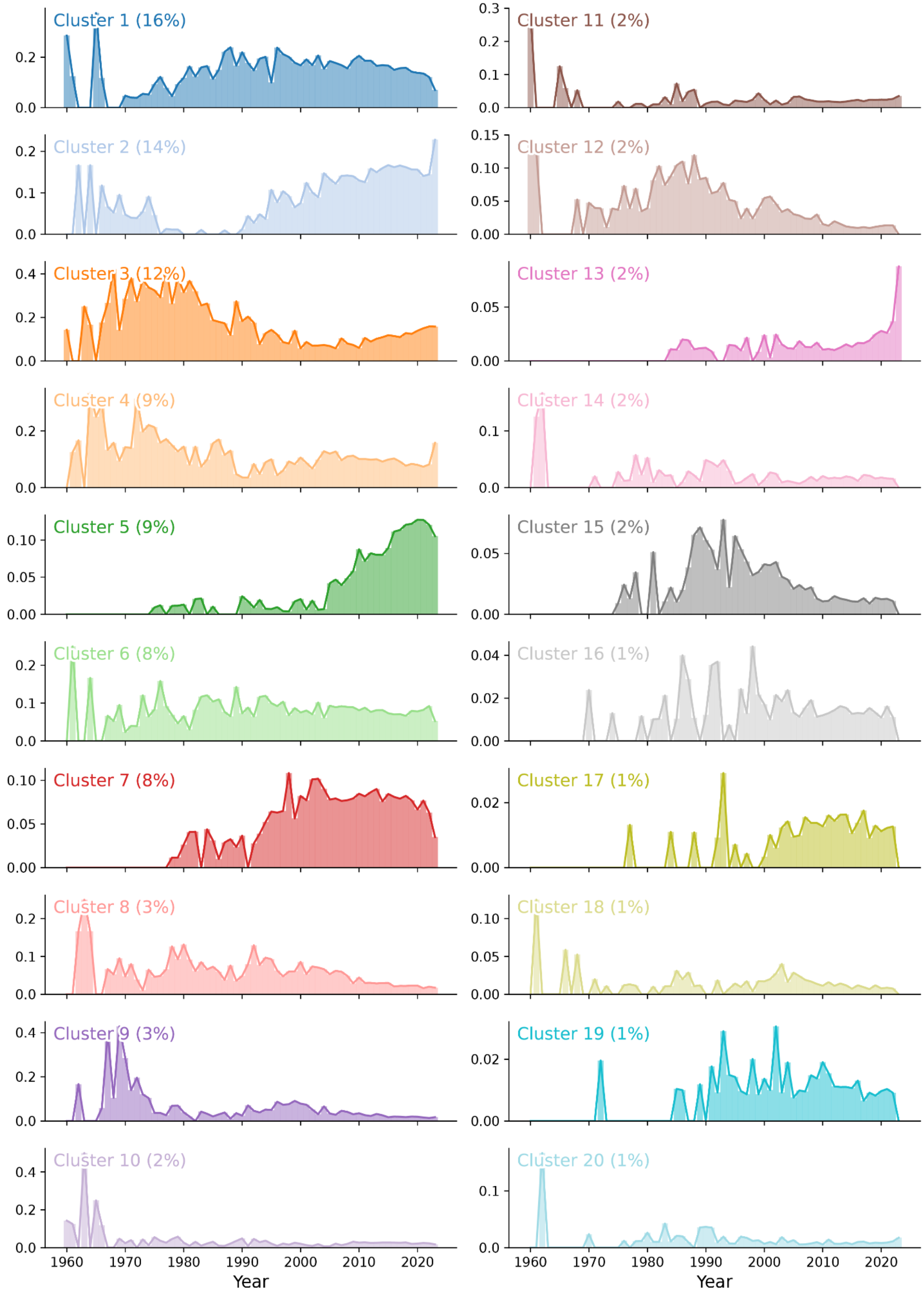

### *Journal Representation of the Field*

Figure 10 presents the topical distribution of articles in core PNPS journals versus the broader corpus. The corpus appears in 9,014 journals across 243 subject categories, indicating a highly dispersed publication landscape. Nearly two-thirds of these journals contain fewer than five PNPS articles, often only one or two.

**Figure 10. Cluster Distribution of Core PNPS Journals**

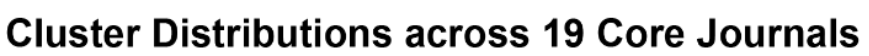
Cluster Distributions across 19 Core Journals

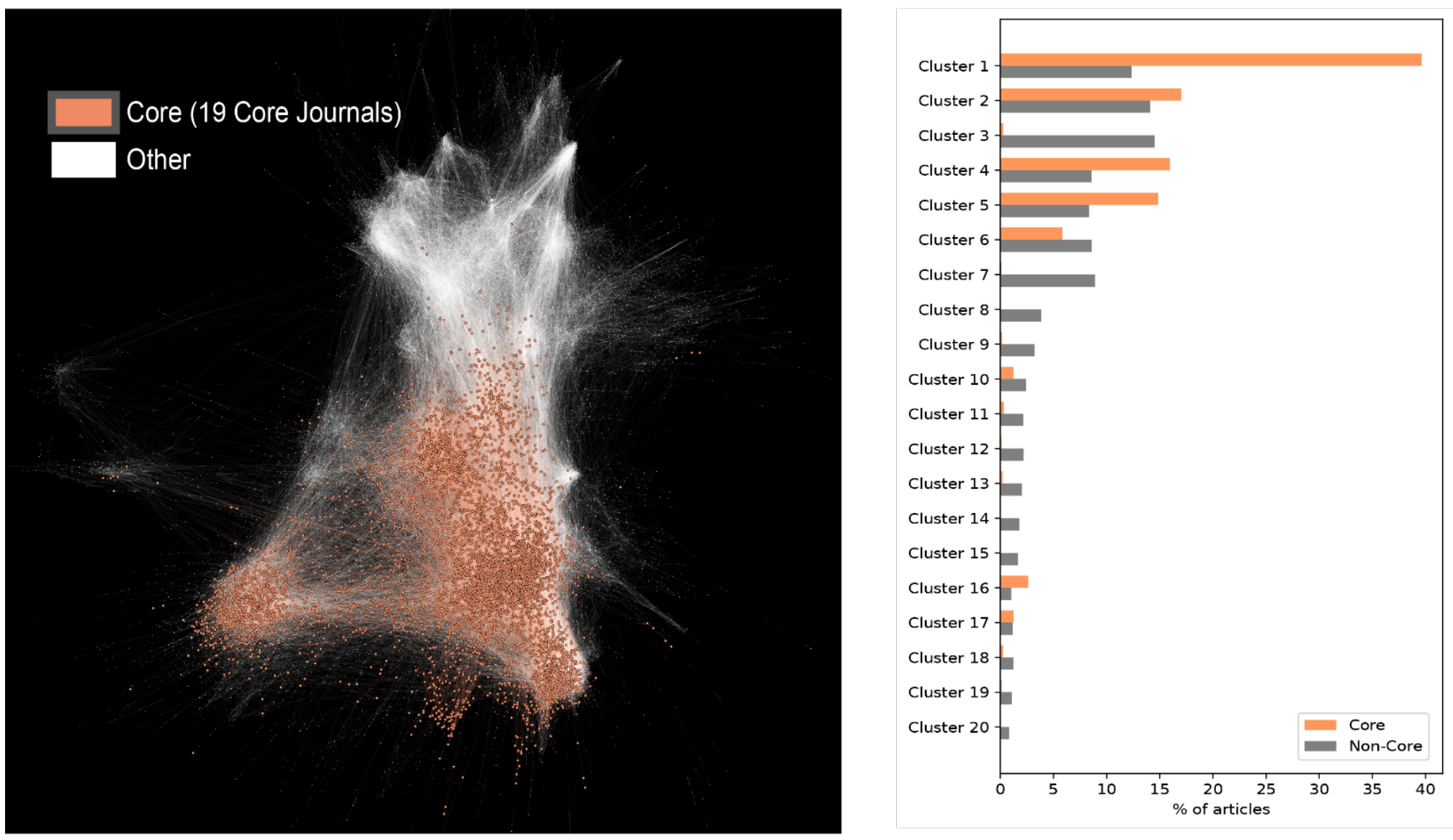

The three flagship journals (i.e., *NVSQ, NML, and VOLUNTAS*) rank among the most prominent outlets publishing PNPS research, accounting for 2,946 articles. Together with sixteen additional mainstream PNPS journals, they account for 6,752 articles. Articles in these journals fall primarily into Clusters 1, 4, 5, and 6, which focus on nonprofit organizations, civic engagement, charitable contributions, and social entrepreneurship. In contrast, several clusters are not covered by the PNPS journals. Research areas such as experimental cooperation, evolutionary perspectives on altruism, global health NGOs, and biological donations are more commonly published in disciplinary outlets in economics, biology, and public health.

### *Underdeveloped and Dispersed Areas*

The network structure also highlights areas that remain structurally dispersed and underdeveloped within the field. As mentioned earlier, Cluster 11 (self-help and mutual aid groups) and Cluster 15 (governance of the commons) are two of the smallest clusters that are most loosely connected with the broader PNPS network. For the former, it is surprising that research on self-help and mutual aid is connected to the rest of the network through only a somewhat unexpected link: faith-based organizations and voluntary social service. This lack of citation ties to the rest of the network is likely due to the fact that our search did not include those keywords related to member-benefit nonprofits (e.g., labor unions, business leagues, trade organizations, fraternal societies, etc.), which are most closely associated with self-help and mutual aid. The latter is also a curious case: the research cluster on the governance of the commons has generated merely 600 articles and is connected with the rest of the network through only one link. By contrast, the seminar work of Elinor Ostrom (1990) and Garrett Hardin (1968) on the subject has each generated over 60,000 Google Scholar citations, which suggests that most of the studies citing this line of work are not in our sample.

Another example of underdeveloped areas is the absence of a distinct cluster focused on legal and public policy issues regarding nonprofits and philanthropy, despite articles from the journal *Nonprofit Policy Forum*. Although regulation, taxation, and policy influence are widely recognized as central to nonprofit activity, related scholarship does not consolidate into a cohesive research community. One possible explanation is that legal and policy-oriented scholarship tends to focus on specific domains, such as taxation and charitable giving or regulatory issues related to nonprofit organizations, which align more closely with existing clusters rather than forming a distinct policy-centered community. Still, this pattern may indicate an opportunity to strengthen the integration and visibility of policy-oriented research within the broader PNPS research agenda.

## DISCUSSION

The emerging interdisciplinary field of PNPS has grown rapidly, yet its intellectual structure remains only partially visible. This study examines publication growth, disciplinary composition, and the citation-based knowledge structure of the field to clarify how PNPS scholarship is organized, where its intellectual core lies, and how its research areas relate to one another. The findings of this study extend and refine existing mapping efforts in PNPS by situating prior insights within a broader citation-based structure. Earlier studies have identified key thematic areas (e.g., volunteering, civic engagement, organizational effectiveness, and economic foundations of the nonprofit sector) and documented the growing cohesion of the field, particularly within core journals. The present analysis partially overlaps with these areas but reveals a different structural configuration: while scholarship on nonprofit organizations remains central to the institutional identity of PNPS, the broader citation structure reveals a far less coherent field dominated by three “power centers” that include not only nonprofit organizations but also social movements and voluntary action (a point we will return to below). Similarly, prior work has shown that PNPS scholarship engages multiple disciplines but exhibits uneven cross-disciplinary connections (e.g., LePere-Schloop & Nesbit, 2022; Walk & Andersson, 2020). The results here build on this insight by identifying where and when such connections occur and/or remain limited, thereby clarifying how adjacent literatures are incorporated into, or remain peripheral to, the broader PNPS knowledge system.

Our study has identified over 60,000 field-related articles appearing in the WoS over the past century, reflecting a substantial expansion of the field. Total citations have increased sharply alongside this growth, while citations per publication have increased more slowly. Existing studies focus on publications in the social sciences and humanities. Although these domains remain dominant, our study finds substantial scholarship in the domains of life sciences and biomedicine, technology, and, to a lesser extent, the physical sciences. Research relevant to prosocial behavior, cooperation, and collective action extends beyond the traditional disciplinary boundaries of PNPS. Greater engagement with scholarship in biomedical and technological domains may yield new perspectives on fundraising, grantmaking, civic engagement, and social impact in contexts increasingly shaped by scientific and technological change.

The citation network analysis provides a structural map of PNPS by identifying 20 major topical clusters and examining their relationships. Rather than simply cataloging topics, the network reveals how these communities connect through patterns of citation exchange. The results indicate that PNPS is dominated by three distinct spheres of influence that do not often speak directly with each other: nonprofit organizations, social movements, and voluntary action. In particular, voluntary action represents an alliance of small and closely linked streams of literature such as charitable contributions, prosocial behaviors, group cooperation, and altruism. This concurrent and yet somewhat siloed presence of three "power centers" can be traced back to the origin of PNPS, when two previously separated fields of nonprofit organizations and voluntary action research merged in the late 1980s. Interestingly, although the early pioneers and advocates for voluntary action research included social movements in their coverage of "voluntary action" (Journal of Voluntary Action Research, 1985), research on social movements has developed its own intellectual autonomy over time and is only weakly and indirectly connected with the other two "power centers." This tripartite co-existence points to opportunities for stronger integration, particularly by reuniting social movement and voluntary action research or by linking organizational research more explicitly with work on giving, contribution, and collective action across contexts.

The publication landscape reflects both the consolidation and dispersion of the field. PNPS-related research appears in more than 9,000 journals across nearly all WoS subject categories, although most journals publish only a small number of relevant articles. The three flagship journals (i.e., *NVSQ, NML, and VOLUNTAS*), together with other PNPS-oriented outlets, primarily focus on research on nonprofit organizations, civic engagement, charitable giving, and related topics, whereas experimental cooperation, evolutionary perspectives on altruism, and health-related nonprofit activity, are more commonly published in disciplinary venues outside PNPS. This pattern suggests that the PNPS journals represent only part of the broader intellectual ecosystem in which PNPS concepts are studied and highlights opportunities for stronger dialogue across adjacent research communities.

Thus, there are many opportunities for cross-disciplinary bridge-building between the centers of power, domains, and weakly connected clusters. Although there are well-known cultural and institutional barriers discouraging such bridge-building, perhaps our findings will persuade scholars and gatekeepers to resist such pressures. If so, journal editors could encourage this through article recruitment, targeted special issues, and guidance to reviewers. Deans and other research administrators could encourage this by supporting the work of interdisciplinary research teams. Research funders can encourage this work through their call-for-proposals guidelines.

Methodologically, this study illustrates how computational bibliometric approaches can enhance the analysis of large and dispersed bodies of scholarship. Citation network analysis provides a systematic way to identify research communities and their relationship, while NLP procedures extract cluster-specific terms. An LLM is used to synthesize these pre-computed terms into concise cluster labels and descriptions for final editing by humans. Similar approaches may be useful for examining the intellectual structure of other interdisciplinary domains (e.g., public administration and policy, business management, and social work). Future research could further assess the reliability of LLM-assisted cluster labeling in social science mapping by developing annotated benchmark datasets and comparing results across LLM models, prompts, and human-coded interpretations.

Several limitations should be acknowledged. First, the citation network stems from the WoS, which underrepresents books, practitioner-oriented outlets, and some smaller journals (Birkle et al., 2020; Singh et al., 2021). This limitation is particularly relevant for PNPS, where books and practitioner publications often play an important role in shaping discourse and informing practice. As a result, the present analysis primarily reflects the structure of journal-based scholarly literature. Advances in digitization techniques may enable future researchers to overcome these limitations by using bibliographic databases such as Google Scholar and Google Books.

A second limitation concerns the delineation of the field itself. The construction of the corpus necessarily depends on keyword selection, and several central terms (e.g., "voluntary," "grant") generate substantial noise in bibliographic searches. This challenge reflects the broader reality that the boundaries of PNPS remain contested and evolving. We address this issue by combining explicit retrieval rules with citation network structure, remaining agnostic on whether some clusters should be regarded as part of PNPS or part of adjacent fields.

Many alternative and related strategies remain possible. For example, leading scholars in PNPS could be surveyed and results used to construct a Guttman scale measuring the centrality of various keywords within the field. Alternatively, the approach of Sinatra et al. (2015) to disciplinary boundaries in physics could be applied here, producing a citation-based approach to determine the probability that an article should be considered part of PNPS. Additional insight could be obtained from complementary networks – networks of coauthorship and mentorship (e.g., thesis committee members or chair), of institutional affiliations, or of article content. These provide structure to the cognitive and social dimensions of the field (Milojević, 2015; Newman, 2004). Future research might also examine the degree of codification (knowledge that can be explicitly represented and categorized using formal languages) within PNPS, which has been linked to the formalization and establishment of a research field (Merton, 1957, p. 507).

In conclusion, this study maps the field of PNPS by combining journal-keyword-based retrieval with citation network analysis. Our mapping functions as a "nomenclature system" (Cheek et al., 2015), helping scholars situate their work, identify gaps, and recognize emerging directions of inquiry. It is our hope that future research will strengthen the coherence of PNPS by aligning its conventional focus with the patterns of connection that structure its wider intellectual connections.

**Notes**

1. The field of philanthropic studies is not only interdisciplinary, but also cross-disciplinary, multidisciplinary, and transdisciplinary (Steinberg, 2004). In this study, we use interdisciplinary as shorthand for all of these.
2. The Leiden algorithm is a stochastic modularity-based community detection method that approximates full optimization. Hence, certain records of publications, usually those located at the boundaries between clusters, are assigned randomly to clusters. Therefore, small variations in cluster sizes occur across optimization rounds. However, the overall structure of the community clusters remains stable and consistent across rounds of analysis (Traag et al., 2019). The algorithm was accessed via Python Library '*leid*enalg' (https://github.com/vtraag/leidenalg). Isolated articles, those neither citing nor cited by other articles, were excluded as they cannot form meaningful clusters.
3. Disparity filters compare results with a null model in which the normalized weights of a node's edges follow a uniform distribution, and retain ties significant at the 0.05 level.
4. The map can be explored interactively at https://pnps2026.surge.sh.
5. Only the top 18 clusters are shown in Figure 6 due to the color scheme designed in the Helios-Web model (Silva, 2023). Studies of visualization show that the maximum number of colors that healthy human eyes can simultaneously differentiate is 20 (Lin et al., 2013). Two colors are used for the background and labels. Therefore, the two smallest clusters are not shown in the visualization.

**Data Availability Statement**

The data are not publicly available due to the proprietary nature of the Web of Science database. Codes are available upon request.
The interactive map of the intellectual structure is available at https://pnps2026.surge.sh, within which readers can search article records using keywords, article titles, journal names, and publication years.

**Acknowledgment**

The authors acknowledge the suggestions provided by Dr. Dwight Burlingame and Dr. Katherine Baderscher throughout the development of the research. This work used JetStream2 at Indiana University through allocation CIS230183 from the Advanced Cyberinfrastructure Coordination Ecosystem: Services & Support (ACCESS) program supported by National Science Foundation grants #2138259, #2138286, #2138307, #2137603, and #2138296. The fourth author thanks the Air Force Office of Scientific Research under Award #FA9550-19-1-0391 for its support. (The funders had no role in study design, data collection and analysis, the decision to publish, or preparation of the manuscript.)

# SUPPLEMENTARY INFORMATION

## A. Curated Journal List (n=19)

1. *Nonprofit and Voluntary Sector Quarterly* (previously named *Journal of Voluntary Action Research*)
2. *Nonprofit Management & Leadership*
3. *VOLUNTAS*
4. *Canadian Journal of Nonprofit and Social Economy Research*
5. *China Nonprofit Review*
6. *Foundation Review*
7. *Journal of Philanthropy and Marketing* (previously named *Journal of Philanthropy*)
8. *Journal of Civil Society*
9. *Journal of Nonprofit Education and Leadership*
10. *Journal of Social Entrepreneurship*
11. *Nonprofit Policy Forum*
12. *Social Enterprise Journal*
13. *Social Movement Studies*
14. *Voluntary Sector Review*
15. *International Review on Public and Nonprofit Marketing*
16. *Journal of Nonprofit & Public Sector Marketing*
17. *Mobilization*
18. *CIRIEC - Espana Revista de Economia Publica Social Y Cooperativa*
19. *Journal of Public and Nonprofit Affairs*

## B. Keyword Curation Process

**Step One: Compile existing keyword lists into an initial list.** Drawing upon existing keyword lists used in bibliometric research on the field (Burlingame, 2004; Gazley & Guo, 2020; LePere-Schloop & Nesbit, 2022; Shier & Handy, 2014; Smith et al., 2006), we created an initial list of keywords by consulting with experts in the field. Many topics are included in some existing work when they define the scope of philanthropic and nonprofit studies, but not others. We resolved the issues by adding and subtracting terms based on our decisions about topics that should be included within the field.

Specifically, the focus of this study is philanthropic actors that are by nature charitable and whose purpose is to provide benefits to the public; therefore, the keywords only include entities classified as 501(c)(3) or 501(c)(4) of the Internal Revenue Code, which constitutes the most common understanding of the nonprofit organization and the nonprofit sector (Hall, 2006). That said, it excludes entities of other 501(c) categories, such as mutual benefit organizations or other types of tax-exempt organizations that focus on the benefit of their members, such as chambers of commerce and trade associations.

The study only focuses on the activity or practice of philanthropic actors, therefore excluding keywords such as "micro-finance" or "cause-related marketing." Further, we include both the formal activities of nonprofit organizations, which have been the primary focus of existing research, and informal aspects of philanthropic activities, which have not received much attention from current research efforts.

**Step Two: Remove keywords to reduce false positives and redundancy.** As a result of the initial list, excessive false positive articles (that have the keywords in their title but are less relevant to the field) and redundant results were identified by the proposed keywords. To eliminate the issues, we exclude the terms based on several criteria.

First, we reject single terms that are of relevance to the field but too broadly defined and most likely infer to the general fields of social science and humanities, such as “social good,” “social change,” “social innovation,” “social impact,” “social responsibility,” “social trust,” “generalized trust,” “social solidarity,” “justice,” “equity,” “equality,” “fairness,” “poverty,” “virtue,” “morality,” “help,” “care,” “share,” “kindness,” “community service,” or “stewardship.”

Second, we reject single terms that are of relevance to the field but are of common use in other existing disciplinary or interdisciplinary fields, such as “social work,” “social service,” “social welfare,” “service learning,” “compassion,” “pro bono,” “quid pro quo,” “civil right movements,” “advocacy,” “refugee,” “human service,” or “humanitarian.”

Third, we reject keywords if they produce no unique, valid results (articles are already included due to other keywords), such as terms that are commonly used to describe methods or tools for investigating certain phenomena of the field, for example, “public good game,” “trust game,” “ultimatum game,” or “dictator game,” or terms that are too narrow or specific in the field, such as “Giving Tuesday” or “fiscal sponsor.”

**Step Three: Adjust variations of keywords to reduce false positives.** We adopt a mixed searching strategy combining singular terms, bi-gram terms, and paired terms to address situations when the singular use of a term returns too many irrelevant records. For example, instead of only searching for “giving,” which returns many irrelevant publications, we searched for “gift giving”, or (“giving”, “pledge”). The form, either bi-gram or paired terms, is determined by the search results of the terms in Google Scholar and the Clarivate-Web of Science (WoS) interfaces. The authors search the first ten pages of Google Scholar and the Clariviate-WoS by searching the focal term and decide which forms of the terms should be included in the final list. To illustrate, “gift giving” is almost always present in this particular sequence, while “giving pledge” is commonly seen in the form of “pledge giving”, “giving pledge”, or “pledge of charitable giving”. Therefore, a paired form is chosen to ensure the search captures the relevant scholarship.

Also, instead of using the vocabulary root as the search term, the author searches all the possible variations of the term when the root solicits non-related items. For example, the root “altruis” for “altruism” and “altruistic,” the search for the root returns records that study actors whose name is “altruis.” However, there are some cases in which a specific variation causes more false positives than the root word; either the particular variation is excluded or included in a bi-gram / paired form. For example, ‘almsgiving’ is included, but ‘almoner’ is not.

In addition, several terms are commonly used to describe the field but simply return many negative positives, such as "independent sector," “voluntary group,” “community interest,” “community trust,” “family trust,” and “settlement house.” These terms are therefore excluded from the final keywords list. Lastly, during the search and analysis process, we identified several keywords that retrieve irrelevant articles, but they do not share any pattern, allowing them to be excluded systematically. We dropped them individually, and they are summarized in the bracket of disallowed keywords.

**Step Four: Evaluate keyword relevance and reliability.** To ensure that curated keywords were effective in retrieving articles truly relevant to philanthropic and nonprofit studies, one author manually reviewed up to the first 100 search results for each keyword in the WoS database. Keywords that yielded no results or returned a substantial proportion of irrelevant or ambiguous articles (approximately above 25% of the total return) were excluded. In addition, certain keywords were systematically excluded when consistent patterns of irrelevant content were observed. For example, while the term "altruistic" was kept, variations such as "altruistic locking" and "altruistic suicide" were excluded due to their frequent association with unrelated topics.

To validate this process, we randomly selected 30 keywords and evaluated the relevance of their top 100 retrieved articles. Three authors independently rated each article based on its title, keywords, journal source, and abstract. We then calculated both raw agreement rates and inter-coder reliability score using Gwet's AC1 (Gwet, 2008) to determine whether each keyword should be kept in the final list. Results are summarized in Table S1. The final list of keywords is presented at the end of this section.

**Table S1. Inter-Coder Reliability and Inclusion of Keywords**

| **Sampled Keyword** | **# records examined** | **# true positive by rater 1** | **# true positive by rater 2** | **# true positive by rater 3** | **Mean % true positive** | **Intercoder Reliability (Gwet's AC1)** | **Included?** |
|---|---|---|---|---|---|---|---|
| grant economies | 0 | 0 | 0 | 0 | NA | NA | NO |
| generous, people | 6 | 5 | 5 | 6 | 88.89% | 0.862 | YES |
| giving, pledge | 16 | 12 | 13 | 13 | 79.17% | 0.938 | YES |
| community organizing | 100 | 99 | 99 | 99 | 99.00% | 1.000 | YES |
| donation, faith | 13 | 13 | 13 | 13 | 100.00% | 1.000 | YES |
| social, reciprocity | 100 | 98 | 100 | 100 | 99.33% | 0.986 | YES |
| nongovernmental organi*ation* | 100 | 99 | 100 | 99 | 99.33% | 0.993 | YES |
| planned giving | 16 | 16 | 16 | 16 | 100.00% | 1.000 | YES |
| cooperative behavio*, human | 36 | 29 | 28 | 30 | 80.56% | 0.838 | YES |
| low-profit limited liability compan* | 2 | 2 | 2 | 2 | 100.00% | 1.000 | YES |
| other-regarding | 100 | 99 | 100 | 98 | 99.00% | 0.980 | YES |
| prosocial* | 100 | 100 | 100 | 100 | 100.00% | 1.000 | YES |
| social economy | 100 | 98 | 100 | 100 | 99.33% | 0.986 | YES |
| trust, civil societ* | 69 | 69 | 69 | 69 | 100.00% | 1.000 | YES |
| nongovernmental institution* | 3 | 3 | 3 | 3 | 100.00% | 1.000 | YES |
| social economies | 27 | 15 | 27 | 16 | 71.60% | 0.459 | NO |
| grassroots group* | 17 | 17 | 17 | 17 | 100.00% | 1.000 | YES |
| eleemosynary organi*ation* | 0 | 0 | 0 | 0 | NA | NA | NO |
| muslim, giving | 33 | 18 | 18 | 16 | 52.53% | 0.839 | NO |
| islam, giving | 18 | 10 | 10 | 8 | 51.85% | 0.852 | NO |

***Curated Keyword List (n=169)***

"philanthropy", "philanthropic", "philanthropist", "misanthropy", "misanthropic", "charity", "charities", "charitable", "benevolence", "benevolent societ*", "altruism", "altruistic", "almsgiving", "gospel of wealth", "gift exchange*", "gift-exchange*", "gifting", "gift giving",

"gift-giving", "collective good*", "public good*", "the commons*", "social economy", "gift economy", "gift economies", "generosity", ("generous, social"), ("generous, people"), ("generous, behavio*"), ("generous, act*"), ("generous, giving"), ("gratitude, giving"), ("gratitude, help*"), ("gratitude, donate"), ("gratitude, donation"), ("gratitude, donating"), ("social capital, civil societ*"), ("trust, civil societ*"), ("communit*, civil societ*"), "the third sector*", "voluntary sector*", "non-distribution constraint*", "nondistribution constraint*", "501c3", "501(c)(3)", "501c4", "501(c)(4)", "civic participation*", "civic engagement*", "social movement*", ("collective action*, social"), ("collective action*, event*"), ("collective action*, group"), ("collective action*, global"), ("collective action*, civil societ*"), "grassroots movement*", "grassroots group*", "grassroots organi*ation*", "grassroots association*", "nonprofit*", "non-profit*", "not for profit*", "not-for-profit*", "nongovernmental organi*ation*", "nongovernmental agenc*", "non-governmental organi*ation*", "non-governmental agenc*", "nongovernmental institution*", "non-governmental institution*", "grantmaking", "grant-making", "grantseeking", "grant-seeking", "endowment foundation*", "endowment fund*", "community foundation*", "family foundation*", "private foundation*", "corporate foundation*", "social enterprise*", "social entrepreneur*", "low-profit limited liability compan*", "b corp", "benefit corporation*", "flexible purpose corporation*", "community interest compan*", "community organizing", "community engagement", "community based organi*tion*", "community-based organi*tion*", "voluntary association*", "voluntary organi*ation*", "selfhelp group*", "self-help group*", "self-help association*", "self-help organi*ation*", "membership association*", "membership organi*ation*", "giving circle*", "giving, pledge", "donor advised fund*", "donor-advised fund*", "corporation giving", "corporate giving", "workplace giving", "impact invest*", "program*related investment*", "volunteering behavio*", "volunteerism", "voluntarism", "voluntaristics", "benefit* of volunteering", "helping other", "helping behavio*", "giving behavio*", "individual giving", "giving money", "giving time", ("donate, money"), ("donate, time, volunteer*"), ("donate, motive*"), ("donate, motivation*"), ("donate, behavio*"), ("donate, decision*"), ("donation, money"), ("donation, time, volunteer*"), ("donation, motive*"), ("donation, motivation*"), ("donation, behavio*"), ("donation, decision*"), "donor fatigue", "anonymous giving", ("giving, diaspora"), "planned giving", ("giving, in-kind"), ("gift, in-kind"), ("giving, wealthy"), "mutual aid", ("social, reciprocity"), "serial reciprocity", ("norm, reciprocity"), ("norms, reciprocity"), ("cooperation, reciprocity"), ("generalized reciprocity, evolution"), ("cooperative behavio*, experiment*"), ("cooperative behavio*, group*"), ("cooperative behavio*, human"), ("cooperative behavio*, animal*"), ("cooperative behavio*, social"), ("cooperation, social"), ("cooperation, human"), ("cooperation, animal*"), "prosocial*", "pro-social*", "empathic behavio*", "empathetic behavio*", "other regarding", "other-regarding", "religious giving", ("giving, religiosity"), ("donation, religio*"), "faith-based giving", "faith giving", ("donation, faith"), ("donation, spiritual*"), "zakat", "tzedakah", "tithe", "waqf"

***Disallowed Keyword Combinations:***
("altruistic","suicide"), ("altruistic locking”),
("altruism","autonomous vehicle"), ("altruism","autonomous driving"),
("charit*","chronic”), ("charit*", "hospital*"),
("charity","mount"), ("charity","salmonella"),
("civil societ*","revolution*"), ("warm*glow","familiarity"),
("doxastic", "volunt*"), ("covergent", "volunt*"), ("theological", "volunt*")

## C. ChatGPT-4o Raw Outputs

***Cluster 1***

**Title [Round 1]**: Dynamics and Challenges in the Nonprofit Sector
**Title [Round 2]**: Dynamics and Strategies in the Nonprofit Sector
**Description [Round 1]**: This cluster explores the multifaceted nature of nonprofit organizations and the sector as a whole. It delves into the financial aspects, management strategies, and accountability mechanisms that define nonprofit organizations. With a focus on the interrelation with government and other partnerships, discussions cover topics such as revenue diversification, the role of professionalism, and the evolving relationships with governmental and non-governmental entities. The cluster also considers the operational challenges and opportunities arising from these dynamics. Various studies address the impact of these interactions on organizational governance, funding, accountability, and the pursuit of mission-oriented goals within the nonprofit sector.
**Description [Round 2]**: This cluster explores the roles, structures, and relationships within the nonprofit sector. It examines the organizational and managerial strategies employed by nonprofit organizations, emphasizing revenue diversification and accountability mechanisms. The interaction between nonprofits, government entities, and the profit sector is also scrutinized to understand partnerships, resource dependence, and governance. Additionally, it investigates the impact of adopting business-like practices and the professionalization within nonprofit organizations, as well as addressing the challenges and opportunities presented by these practices.
**Top 100 grams (The number in parentheses is the value of relative frequency.)**:
nonprofit(0.39), sector(0.20), organization(0.20), nonprofit organization(0.17), profit(0.16), non profit(0.09), organizational(0.09), nonprofit sector(0.09), service(0.08), organizations(0.07), management(0.07), funding(0.07), ngos(0.07), accountability(0.06), government(0.06), profit organization(0.05), board(0.05), governance(0.05), ngo(0.05), financial(0.04), governmental(0.04), non governmental(0.04), civil society(0.04), revenue(0.04), profit sector(0.04), npos(0.04), nonprofit organizations(0.04), hospital(0.04), organisation(0.03), mission(0.03), partnership(0.03), human service(0.03), foundation(0.03), agency(0.03), social service(0.03), executive(0.03), nongovernmental(0.03), organization ngos(0.03), governmental organization(0.03), fund(0.03), delivery(0.03), civil(0.03), nonprofit management(0.03), nonprofits(0.03), director(0.03), npo(0.02), nongovernmental organization(0.02), grant(0.02), voluntary sector(0.02), performance(0.02), profit organisation(0.02), manager(0.02), organization npos(0.02), accounting(0.02), provider(0.02), service delivery(0.02), staff(0.02), funder(0.02), service organization(0.02), collaboration(0.02), capacity(0.02), advocacy(0.02), reporting(0.02), government nonprofit(0.02), profit hospital(0.02), nonprofit profit(0.02), administrative(0.02), public nonprofit(0.02), philanthropic(0.02), profit nonprofit(0.02), role nonprofit(0.02), service nonprofit(0.02), nonprofit hospital(0.02), transparency(0.02), management nonprofit(0.01), board member(0.01), nonprofit human(0.01), study nonprofit(0.01), public sector(0.01), nonprofit board(0.01), profit profit(0.01), public service(0.01), nonprofit manager(0.01), organization nonprofit(0.01), government funding(0.01), society organization(0.01), federal(0.01), exempt(0.01), contract(0.01), foundations(0.01), board director(0.01), capacity building(0.01), stakeholder(0.01), cso(0.01), resource dependence(0.01), public management(0.01), client(0.01), disclosure(0.01), interorganizational(0.01), contracting(0.01),
**Top 10 most central articles**:

1. The Role Of Nonprofit Enterprise
2. Diversification Of Revenue Strategies: Evolving Resource Dependence In Nonprofit Organizations
3. Alternative Models Of Government-Nonprofit Sector Relations: Theoretical And International Perspectives
4. The Rationalization Of Charity: The Influences Of Professionalism In The Nonprofit Sector
5. The Sociology Of Nonprofit Organizations And Sectors
6. Too Close For Comfort? The Impact Of Official Aid On Nongovernmental Organizations
7. Accountability In Practice: Mechanisms For Ngos
8. The Purpose (And Perils) Of Government-Nonprofit Partnership
9. Being "Business-Like" In A Nonprofit Organization: A Grounded And Inductive Typology
10. Privatizing The Welfare State: Nonprofit Community-Based Organizations As Political Actors

***Cluster 2***

**Title [Round 1]**: Dynamics of Social and Political Movements
**Title [Round 2]**: Dynamics of Social and Political Movements
**Description [Round 1]**: This cluster examines the intricate dynamics, organization, and impact of social and political movements. It highlights the significance of social movements as mechanisms for political protest and activism, often mobilized around various causes and collective actions. The research delves into the underlying theories of movement organization and mobilization, demonstrating how activists and political entities engage in protest, struggle, and collective action across different mediums and spaces. The studies also explore the framing and discourse surrounding movements, along with the role of tactics and radical approaches in contentious politics. The sociological perspective on these movements is essential for understanding their role in shaping contemporary political discourse and their impact on social justice, democratic processes, and global movements. Through empirical and theoretical analysis, the cluster provides insights into the evolution, opportunities, and constraints of protest movements, with a particular focus on social identity, media interaction, and transnational implications.
**Description [Round 2]**: This cluster examines the multifaceted phenomena of social and political movements. It delves into various aspects such as the mechanics of mobilization, the role of activism, and the framing of protest and resistance. Key areas of focus include the exploration of collective identity, political opportunities, and the intersection of media and social movements. The cluster also highlights the role of transnational networks and democratic discourse in shaping movements, resistance strategies, and grassroots actions. Additionally, it discusses contemporary challenges and the transformation of social movements within the political landscape, offering insights from ethnographic studies and analyses of digital and media influence on contentious politics.
**Top 100 grams (The number in parentheses is the value of relative frequency.)**: movement(0.78), social movement(0.76), political(0.30), protest(0.24), activist(0.20), politic(0.17), activism(0.17), mobilization(0.15), collective(0.14), action(0.10), article(0.10), movements(0.10), argue(0.09), struggle(0.08), anti(0.08), social movements(0.08), collective action(0.08), frame(0.08), medium(0.08), actor(0.07), space(0.07), resistance(0.07), mobilize(0.06), identity(0.06), movement organization(0.06), transnational(0.05), democracy(0.05), power(0.05), movement theory(0.05), movement social(0.05), discourse(0.05), contemporary(0.05), framing(0.05), tactic(0.05), contentious(0.05), radical(0.05), political

opportunity(0.04), campaign(0.04), justice(0.04), collective identity(0.04), claim(0.04), event(0.04), union(0.04), social medium(0.04), occupy(0.04), grassroots(0.04), contention(0.04), street(0.04), neoliberal(0.04), civil(0.04), repertoire(0.04), democratic(0.04), violence(0.04), draw(0.04), organize(0.04), mobilisation(0.03), ethnographic(0.03), opposition(0.03), coalition(0.03), solidarity(0.03), feminist(0.03), protester(0.03), social change(0.03), transformation(0.03), scholar(0.03), sociology(0.03), global(0.03), demonstration(0.03), movement study(0.03), scholarship(0.03), construction(0.03), discursive(0.03), media(0.03), movement political(0.03), urban(0.03), new social(0.03), repression(0.03), shape(0.03), emergence(0.03), opportunity structure(0.03), latin(0.03), counter(0.03), alliance(0.03), protest movement(0.03), politics(0.03), crisis(0.03), party(0.03), mass(0.03), rights(0.02), protest social(0.02), digital(0.02), city(0.02), protest event(0.02), network(0.02), narrative(0.02), conflict(0.02), latin america(0.02), human right(0.02), contentious politic(0.02), ideological(0.02),

**Titles of the top 10 most central articles**:

1. Protest And Political Opportunities
2. The Emotions Of Protest: Affective And Reactive Emotions In And Around Social Movements
3. Social Movement Organizations - Growth, Decay And Change
4. An insider's critique of the social movement framing perspective
5. The use of newspaper data in the study of collective action
6. Debunking Spontaneity: Spain's 15-M/Indignados as Autonomous Movement
7. Rethinking Prefiguration: Alternatives, Micropolitics and Goals in Social Movements
8. Collective Identity in Social Movements: Central Concepts and Debates
9. Making the News: Movement Organizations, Media Attention, and the Public Agenda
10. Internet And Social Movement Action Repertoires Opportunities And Limitations

***Cluster 3***

**Title [Round 1]**: Development and Influences on Prosocial Behavior

**Title [Round 2]**: Empathy, Prosocial Behavior, and Developmental Dynamics

**Description [Round 1]**: This cluster examines the development and influences of prosocial behavior, particularly among children and adolescents. It discusses how factors like empathy, age, and emotional awareness contribute to the emergence of prosocial tendencies such as helping and altruism. It looks at self-report measures and parent or teacher observations to analyze prosocial and antisocial behaviors, both distinct and interconnected during developmental stages from childhood to adolescence. Key themes include the role of empathy, social influences, cognitive development, and the impact of social exclusion or class on such behaviors.

**Description [Round 2]**: This cluster explores the relationship between empathy and prosocial behaviors across different ages, focusing on children and adolescents. Key areas of investigation include how these behaviors manifest and develop, the factors influencing them, such as emotional and social dynamics, and their measurement methods. It also examines how various developmental stages, gender differences, and parental influences impact prosocial tendencies and related behaviors like altruism and aggression. Studies in this cluster aim to understand both individual differences and social influences affecting prosocial tendencies, applying self-reports and behavioral assessments across different contexts and age groups.

**Top 100 grams (The number in parentheses is the value of relative frequency.)**:

prosocial(0.55), behavior(0.38), prosocial behavior(0.38), child(0.20), age(0.17), empathy(0.14), adolescent(0.12), participant(0.11), prosociality(0.09), emotional(0.09), predict(0.09), emotion(0.08), year(0.07), year old(0.07), peer(0.07), old(0.07), prosocial behaviour(0.07), girl(0.07), age year(0.07), empathic(0.06), complete(0.06), help behavior(0.06), boy(0.06), aggression(0.06), social behavior(0.06), antisocial(0.06), report(0.06), self report(0.06), school(0.06), cognitive(0.06), measure(0.06), parent(0.06), result show(0.05), adolescence(0.05), current study(0.05), childhood(0.05), present study(0.05), young(0.05), aggressive(0.05), teacher(0.05), questionnaire(0.05), task(0.05), positive(0.05), study examine(0.05), moral(0.05), helping(0.05), altruism(0.05), children(0.05), tendency(0.05), correlate(0.05), altruistic(0.04), mediate(0.04), child prosocial(0.04), adult(0.04), behavioral(0.04), associate(0.04), show(0.04), self(0.04), pro social(0.04), student(0.04), female(0.04), psychological(0.04), behaviors(0.04), young child(0.04), personality(0.04), developmental(0.04), grade(0.04), early(0.04), interpersonal(0.04), child age(0.04), old child(0.03), positively(0.03), male(0.03), preschool(0.03), behaviour(0.03), parental(0.03), mother(0.03), distress(0.03), sample(0.03), longitudinal(0.03), neural(0.03), individual difference(0.03), altruistic behavior(0.03), pro(0.03), cognition(0.03), behavior child(0.03), moderate(0.03), score(0.03), adolescents(0.03), antisocial behavior(0.03), negatively(0.03), empathic concern(0.03), behavior prosocial(0.03), friend(0.03), empathy prosocial(0.03), trait(0.03), parenting(0.03), prosocial antisocial(0.03), negative(0.03), increase prosocial(0.03),

**Titles of the top 10 most central articles**:

1. The Relation Of Empathy To Pro-Social And Related Behaviors
2. The Development Of A Measure Of Prosocial Behaviors For Late Adolescents
3. The Altruistic Personality And The Self-Report Altruism Scale
4. Reinterpreting The Empathy-Altruism Relationship: When One Into One Equals Oneness
5. Having Less, Giving More: The Influence Of Social Class On Prosocial Behavior
6. Prosocial Spending And Well-Being: Cross-Cultural Evidence For A Psychological Universal
7. Social Exclusion Decreases Prosocial Behavior
8. Altruistic Helping In Human Infants And Young Chimpanzees
9. Is Empathic Emotion A Source Of Altruistic Motivation
10. Sociocognitive And Behavioral Correlates Of A Measure Of Prosocial Tendencies For Adolescents

***Cluster 4***

**Title [Round 1]**: Civic Engagement and Volunteerism

**Title [Round 2]**: Understanding Civic Engagement and Participation

**Description [Round 1]**: This cluster explores the multifaceted concept of civic engagement, focusing on the roles of volunteering and participation within communities. It considers the motivations, definitions, and implications of volunteer work and civic participation in both youth and adults. The cluster addresses how civic involvement can impact social capital and political participation, examining voluntary association memberships across different countries and cultural contexts. It reflects on how civic engagement affects community organizing, political engagement, and the development of civic skills among different demographics, including students and immigrants.

**Description [Round 2]**: This cluster explores the multifaceted concept of civic engagement, focusing on its definition, motivations, and implications within various contexts. It examines the role of volunteering and participation in building social capital, fostering community involvement, and influencing political engagement. The research also delves into factors that affect sustained volunteerism, such as dispositional influences and organizational support, and the impact of voluntary association membership across different countries. Additionally, the cluster addresses the transition from youth to adulthood in terms of civic engagement, highlighting how early involvement in civic activities can shape future political participation and community involvement.

**Top 100 grams (The number in parentheses is the value of relative frequency.)**:
civic(0.39), civic engagement(0.32), engagement(0.31), volunteer(0.28), volunteering(0.20), participation(0.18), volunteerism(0.14), youth(0.09), civic participation(0.09), voluntary(0.08), social capital(0.07), survey(0.06), voluntary association(0.06), adult(0.06), community(0.05), involvement(0.05), political participation(0.05), student(0.05), young(0.05), citizen(0.04), citizenship(0.04), young people(0.04), capital(0.04), association(0.04), membership(0.04), youth civic(0.04), old adult(0.03), participation civic(0.03), education(0.03), satisfaction(0.03), voluntary organization(0.03), volunteers(0.03), volunteer work(0.03), participate(0.03), service learning(0.03), motivation volunteer(0.03), political engagement(0.03), associations(0.02), college(0.02), community service(0.02), neighborhood(0.02), civic political(0.02), engagement social(0.02), level civic(0.02), civically(0.02), engagement political(0.02), volunteer activity(0.02), recruitment(0.02), retention(0.02), volunteer management(0.02), community organizing(0.02), voluntary associations(0.02), engagement civic(0.02), active(0.02), immigrant(0.02), civic activity(0.02), life(0.02), civic education(0.02), volunteer motivation(0.02), learning(0.02), engagement community(0.02), form civic(0.02), volunteer volunteer(0.02), volunteer experience(0.02), youth development(0.02), college student(0.02), survey datum(0.02), high school(0.02), socialization(0.02), associational(0.02), promote civic(0.02), school(0.02), old(0.02), organizing(0.02), participation voluntary(0.02), skill(0.02), efficacy(0.02), community organize(0.02), education civic(0.02), civic life(0.02), engaged(0.02), course(0.01), ethnic(0.01), participation political(0.01), civically engage(0.01), formal volunteering(0.01), aging(0.01), association membership(0.01), student civic(0.01), likely volunteer(0.01), volunteer satisfaction(0.01), engagement youth(0.01), young adult(0.01), participation community(0.01), formal(0.01), community engagement(0.01), adulthood(0.01), volunteer organization(0.01), voting(0.01), capital civic(0.01),

**Titles of the top 10 most central articles**:
1. What Do We Mean By "Civic Engagement"?
2. Dispositional And Organizational Influences On Sustained Volunteerism: An Interactionist Perspective
3. The Motivations To Volunteer: Theoretical And Practical Considerations
4. Defining Who Is A Volunteer: Conceptual And Empirical Considerations
5. Civic Engagement And The Transition To Adulthood
6. Voluntary Association Membership In 15 Countries - A Comparative-Analysis
7. Nations Of Joiners: Explaining Voluntary Association Membership In Democratic Societies
8. Does Participation In Voluntary Associations Contribute To Social Capital? The Impact Of Intensity, Scope, And Type
9. Volunteerism: Social Issues Perspectives And Social Policy Implications
10. Bowling Young: How Youth Voluntary Associations Influence Adult Political Participation

***Cluster 5***

**Title [Round 1]**: Social Entrepreneurship and Hybrid Organizations: Research and Development
**Title [Round 2]**: Social Entrepreneurship and Enterprise Dynamics

**Description [Round 1]**: This cluster encompasses the study of social entrepreneurship and its impact on enterprises and the economy. It explores the characteristics and motives of entrepreneurs engaged in social enterprises, as well as the methodologies they employ to create social value. The research delves into the beneficial role of hybrid organizations that incorporate both commercial and social objectives. Emphasis is placed on the unique challenges and opportunities in implementing sustainable development and innovation in the social economy. By examining case studies and typologies of social entrepreneurs, the cluster provides insights into the multifaceted nature of social enterprises, their development as a field, and implications for future research.
**Description [Round 2]**: This cluster explores the intersection of entrepreneurship and social impact, focusing on enterprises that prioritize both economic and social objectives. It discusses the role of entrepreneurs in shaping businesses that not only aim for economic gains but also seek to address social challenges. The keywords emphasize the importance of innovation, sustainability, and value creation within the context of social entrepreneurship. The sample titles suggest a critical evaluation of existing theories, the hybrid nature of social enterprises, and the different motives and ethical considerations that influence social entrepreneurs. This body of work also investigates the convergence and divergence of social entrepreneurship concepts across different geographic locations, providing a comprehensive understanding of the diverse landscape of social enterprises.
**Top 100 grams (The number in parentheses is the value of relative frequency.)**:
enterprise(0.48), entrepreneurship(0.46), social enterprise(0.45), social entrepreneurship(0.42), entrepreneur(0.25), business(0.22), entrepreneurial(0.20), social entrepreneur(0.19), innovation(0.16), economy(0.12), methodology(0.12), entrepreneurship social(0.11), originality(0.11), purpose(0.10), design methodology(0.10), sustainable(0.10), social economy(0.10), venture(0.10), enterprise social(0.10), value(0.10), social innovation(0.09), creation(0.09), originality value(0.09), methodology approach(0.09), social entrepreneurial(0.09), economic(0.08), case study(0.07), approach(0.07), sustainability(0.07), market(0.07), enterprises(0.07), innovative(0.07), social enterprises(0.07), social impact(0.07), purpose paper(0.06), social value(0.06), mission(0.06), findings(0.06), social economic(0.06), aim(0.06), business model(0.06), development social(0.06), commercial(0.06), hybrid(0.05), stakeholder(0.05), create(0.05), employment(0.05), development(0.05), company(0.05), social problem(0.05), innovation social(0.05), value creation(0.04), research limitation(0.04), social mission(0.04), paper aim(0.04), limitation implication(0.04), economic social(0.04), limitation(0.04), design(0.04), sustainable development(0.04), qualitative(0.04), social business(0.03), develop(0.03), model social(0.03), social venture(0.03), social environmental(0.03), role social(0.03), study social(0.03), create social(0.03), start(0.03), objective(0.03), achieve(0.03), hybrid organization(0.03), economy social(0.03), solution(0.03), financial(0.03), ses(0.03), entrepreneurial intention(0.03), initiative(0.03), institutional(0.03), field social(0.03), entrepreneurial activity(0.03), concept social(0.03), ecosystem(0.03), creation social(0.03), impact social(0.03), approach paper(0.03), paper(0.03), entity(0.03), integration(0.03), entrepreneur social(0.03), capability(0.03), value paper(0.03), conceptual(0.02), investor(0.02), work integration(0.02), logic(0.02), entrepreneurship research(0.02), concept(0.02), depth(0.02),
**Titles of the top 10 most central articles**:
1. Social Entrepreneurship Research: A Source Of Explanation, Prediction, And Delight
2. A Typology Of Social Entrepreneurs: Motives, Search Processes And Ethical Challenges

3. Social Enterprises As Hybrid Organizations: A Review And Research Agenda
4. Conceptions Of Social Enterprise And Social Entrepreneurship In Europe And The United States: Convergences And Divergences
5. Social Entrepreneurship: A Critical Review Of The Concept
6. Social Entrepreneurship: Why We Don't Need A New Theory And How We Move Forward From Here
7. Research In Social Entrepreneurship: Past Contributions And Future Opportunities
8. Advancing Research On Hybrid Organizing - Insights From The Study Of Social Enterprises
9. Social Entrepreneurship: A Critique and Future Directions
10. Enterprising nonprofits

***Cluster 6***
**Title [Round 1]**: Charitable Giving and Altruism
**Title [Round 2]**: Understanding Charitable Behavior: Motivations and Dynamics
**Description [Round 1]**: This cluster explores the motivations, behaviors, and mechanisms underlying charitable giving, including donations, fundraising, and donor behavior. It focuses on the concepts of warm-glow giving, impure altruism, and the private provision of public goods. The keywords indicate a strong interest in the experimental approach, including field and laboratory experiments, to analyze how incentives, social norms, and image motivations affect charitable contributions. The titles reflect theoretical and applied perspectives on how monetary incentives, prestige, and social pressures influence the act of giving to charities and the provision of public goods.
**Description [Round 2]**: This cluster explores the complex motivations and behaviors behind charitable giving and donations. It focuses on the role of altruism, both pure and impure, in influencing donor behavior and the decision-making process. Concepts like 'warm-glow' giving, where donors derive emotional satisfaction from donating, are examined alongside the impact of financial incentives and social pressures. The studies included probe into how charitable actions are affected by various factors such as income, tax legislation, marketing strategies, and fundraising techniques. They also consider the dynamics of private provision of public goods, willingness to donate, and the effectiveness of different organizational appeals. Through field and laboratory experiments, the cluster analyzes how donors respond to solicitation efforts, branding, and advertising campaigns, as well as the implications of donation behaviors on public and private goods provisions.
**Top 100 grams (The number in parentheses is the value of relative frequency.)**:
charitable(0.28), donation(0.28), giving(0.21), charity(0.19), charitable giving(0.19), donor(0.16), donate(0.14), experiment(0.09), fundraising(0.06), give(0.06), charitable donation(0.06), field experiment(0.06), money(0.05), appeal(0.05), incentive(0.04), marketing(0.04), consumer(0.04), crowd(0.04), elsevi right(0.04), preference(0.03), charitable contribution(0.03), warm(0.03), price(0.03), glow(0.03), warm glow(0.03), contribution(0.03), increase donation(0.03), tax(0.03), public good(0.03), charitable organization(0.03), cause(0.03), monetary(0.03), elsevi(0.03), provision public(0.03), intention(0.03), private provision(0.03), willingness(0.03), donation behavior(0.02), income(0.02), fundraise(0.02), potential donor(0.02), impure(0.02), brand(0.02), solicitation(0.02), crowdfunding(0.02), blood(0.02), advertising(0.02), donation charity(0.02), pure(0.02), blood donation(0.02), intention donate(0.02), amount(0.02), fundraiser(0.02), donation intention(0.02), campaign(0.02), give behavior(0.02), charitable behavior(0.02), request(0.02), crowdfunde(0.02), donate money(0.02), charitable give(0.02), generosity(0.02), image(0.02), good provision(0.02), charitable

cause(0.02), estimate(0.02), willingness donate(0.02), donate charity(0.02), donations(0.02), charity donation(0.01), household(0.01), crowding(0.01), blood donor(0.01), message(0.01), donation decision(0.01), deduction(0.01), elasticity(0.01), voluntary contribution(0.01), lottery(0.01), dictator(0.01), purchase(0.01), increase charitable(0.01), solicit(0.01), pure public(0.01), payment(0.01), provision(0.01), voluntary provision(0.01), lab(0.01), effect charitable(0.01), vs.(0.01), matching(0.01), dictator game(0.01), social norm(0.01), equilibrium(0.01), alumnus(0.01), laboratory(0.01), laboratory experiment(0.01), advertisement(0.01), impure public(0.01), tax incentive(0.01),
**Titles of the top 10 most central articles**:
1. Impure Altruism And Donations To Public-Goods - A Theory Of Warm-Glow Giving
2. Giving With Impure Altruism - Applications To Charity And Ricardian Equivalence
3. Incentives And Prosocial Behavior
4. Privately Provided Public-Goods In A Large Economy - The Limits Of Altruism
5. Doing Good Or Doing Well? Image Motivation And Monetary Incentives In Behaving Prosocially
6. Global Public Goods: A Survey
7. The Private Provision Of A Public Good Is Independent Of The Distribution Of Income
8. A Signaling Explanation For Charity
9. What Do Donations Buy? A Model Of Philanthropy Based On Prestige And Warm Glow
10. Testing For Altruism And Social Pressure In Charitable Giving

***Cluster 7***
**Title [Round 1]**: Mechanisms of Cooperation and Punishment in Public Goods Games
**Title [Round 2]**: Cooperation and Punishment in Public Goods Games
**Description [Round 1]**: This cluster explores the dynamics of cooperation in public goods games, focusing on the concepts of altruistic punishment, conditional cooperation, and the role of reputation in sustaining cooperative behavior. Through experimental and evolutionary game theory models, the studies examine how public goods dilemmas, such as free riding and social preferences, affect collective contributions and individual decision making. Laboratory experiments provide evidence on mechanisms that promote cooperation, including costly punishment and reputation management, and explore how social norms and diversity influence cooperative outcomes.
**Description [Round 2]**: This cluster examines the dynamics of cooperation and punishment within the context of public goods games. It explores how these concepts are studied through experimental and game-theoretical frameworks. The focus is on understanding how cooperative behavior emerges, the role of altruistic punishment, and factors like social dilemmas, reputation, and social preferences that influence individual contributions to public goods. The titles in this cluster highlight experimental evidence and theoretical insights into how cooperative behavior can be sustained or disrupted by free riding, evolutionary mechanisms, and various incentives.
**Top 100 grams (The number in parentheses is the value of relative frequency.)**:
game(0.50), public good(0.41), cooperation(0.38), good(0.34), good game(0.31), experiment(0.29), experimental(0.17), dilemma(0.16), right reserve(0.16), reserve(0.16), punishment(0.15), evolutionary(0.14), public(0.14), player(0.12), cooperative(0.12), payoff(0.11), social dilemma(0.11), evolution(0.11), cooperator(0.11), contribution(0.10), elsevi(0.10), right(0.09), free(0.09), preference(0.09), reciprocity(0.09), elsevi right(0.09), subject(0.09), punish(0.08), individual(0.08), laboratory(0.08), evolution cooperation(0.08), game theory(0.08), elsevier(0.08), good experiment(0.08), cooperation public(0.07),

simulation(0.07), cooperate(0.07), defector(0.07), mechanism(0.07), evolutionary game(0.07), promote cooperation(0.06), costly(0.06), laboratory experiment(0.06), threshold(0.06), public goods(0.06), group member(0.05), goods(0.05), equilibrium(0.05), cooperative behavior(0.05), dictator(0.05), voluntary contribution(0.05), contribution public(0.05), agent(0.05), free rider(0.05), dictator game(0.05), rider(0.05), level cooperation(0.05), reputation(0.04), repeat(0.04), social preference(0.04), incentive(0.04), selfish(0.04), spatial public(0.04), heterogeneous(0.04), decision(0.04), conditional(0.04), parameter(0.04), prisoner dilemma(0.04), spatial(0.04), prisoner(0.04), experimentally(0.04), elsevier inc.(0.04), cost(0.04), pool(0.04), free riding(0.04), riding(0.04), fairness(0.04), observe(0.04), altruistic punishment(0.04), selection(0.04), experimental evidence(0.04), defection(0.04), sanction(0.04), inc.(0.03), average(0.03), treatment(0.03), round(0.03), situation(0.03), probability(0.03), cooperation level(0.03), norm(0.03), choose(0.03), good provision(0.03), pgg(0.03), experimental economic(0.03), social norm(0.03), common pool(0.03), stable(0.03), free ride(0.03), conditional cooperation(0.03),

**Titles of the top 10 most central articles**:

1. Altruistic punishment in humans
2. Cooperation and punishment in public goods experiments
3. Are people conditionally cooperative? Evidence from a public goods experiment
4. Sustaining cooperation in laboratory public goods experiments: a selective survey of the literature
5. Reputation helps solve the 'tragedy of the commons'
6. Social Preferences, Beliefs, and the Dynamics of Free Riding in Public Goods Experiments
7. Social diversity promotes the emergence of cooperation in public goods games
8. Social distance and other-regarding behavior in dictator games
9. Personality and Prosocial Behavior: A Theoretical Framework and Meta-Analysis
10. The evolution of altruistic punishment

***Cluster 8***

**Title [Round 1]**: The Evolutionary Dynamics of Altruism and Cooperation

**Title [Round 2]**: Evolutionary Dynamics of Altruism and Cooperation

**Description [Round 1]**: This cluster explores the complex relationships between altruism, evolution, and selection processes across various species. It investigates how evolutionary mechanisms, such as kin selection and reciprocal altruism, influence behaviors within populations, impacting fitness, reproductive success, and cooperation amongst individuals. The role of genetic factors, relatedness, and environmental interactions in shaping altruistic and cooperative traits is thoroughly examined. Primary studies also focus on specific animal behaviors, cooperative breeding, and the challenges posed by competition, such as the tragedy of the commons. The integration of mathematical and biological models helps to understand the conditions under which these traits evolve and stabilize in both simple and complex ecological settings.

**Description [Round 2]**: This cluster explores the intricate dynamics of altruism and cooperation in evolutionary biology. Central to this discussion is the concept of altruism, where actions that benefit others incur a cost to the individual. Topics like kin selection, inclusive fitness, and the evolution of cooperative behaviors in animals are examined, delving into how these behaviors impact evolutionary fitness and reproductive success. The notion of group selection is also addressed, highlighting the evolutionary advantages of cooperative breeding and the role of genetic and environmental factors in the evolution of altruistic behaviors. Various species, including primates, birds, and bacteria, are analyzed to understand how altruism evolves in different ecological contexts, considering factors like kinship, relatedness, and reciprocity.
**Top 100 grams (The number in parentheses is the value of relative frequency.)**: altruism(0.40), evolution(0.35), selection(0.28), evolutionary(0.22), fitness(0.22), kin(0.21), specie(0.19), cooperation(0.18), kin selection(0.16), population(0.16), cooperative(0.15), animal(0.13), evolve(0.12), altruistic(0.12), genetic(0.12), behaviour(0.11), reproductive(0.10), relatedness(0.10), reciprocal(0.10), trait(0.10), reciprocal altruism(0.09), interaction(0.09), food(0.08), cell(0.08), evolution altruism(0.08), breeding(0.08), offspring(0.08), competition(0.07), inclusive fitness(0.07), gene(0.07), primate(0.07), natural(0.07), group selection(0.07), benefit(0.07), organism(0.07), conspecific(0.06), hamilton(0.06), cost(0.06), helper(0.06), partner(0.06), male(0.06), breed(0.06), relative(0.06), produce(0.06), cooperative breeding(0.06), reproduction(0.06), bacteria(0.06), individual(0.06), simple(0.06), tragedy common(0.06), occur(0.06), tragedy(0.05), selfish(0.05), social evolution(0.05), inclusive(0.05), rat(0.05), female(0.05), dispersal(0.05), colony(0.05), cooperatively(0.05), stable(0.05), simulation(0.05), biological(0.05), biology(0.05), pair(0.05), wild(0.05), insect(0.05), evolution cooperation(0.05), frequency(0.05), survival(0.05), microbial(0.05), altruist(0.05), nest(0.05), ecological(0.05), natural selection(0.05), hamilton rule(0.05), cooperatively breed(0.05), bacterial(0.05), breeder(0.04), favour(0.04), chimpanzee(0.04), mutant(0.04), cooperate(0.04), phenotype(0.04), favor(0.04), costly(0.04), sex(0.04), condition(0.04), predator(0.04), evolutionarily(0.04), reciprocity(0.04), size(0.04), cheat(0.04), species(0.04), bird(0.04), help behaviour(0.04), kinship(0.04), sociality(0.04), plant(0.04), human(0.04),
**Titles of the top 10 most central articles**:
1. Evolution Of Reciprocal Altruism
2. Tragedy Of The Commons As A Result Of Root Competition
3. Public Goods In Relation To Competition, Cooperation, And Spite
4. The Evolution Of Cooperative Breeding Through Group Augmentation
5. Altruism In Viscous Populations - An Inclusive Fitness Model
6. Viscous Medium Promotes Cooperation In The Pathogenic Bacterium Pseudomonas Aeruginosa
7. Explaining The Sociobiology Of Pyoverdin Producing Pseudomonas: A Comment On Zhang And Rainey (2013)
8. Diminishing Returns In Social Evolution: The Not-So-Tragic Commons
9. Perspective: Repression Of Competition And The Evolution Of Cooperation
10. Empathy And Pro-Social Behavior In Rats

***Cluster 9***
**Title [Round 1]**: Public Goods and Their Provision in Different Contexts
**Title [Round 2]**: Provision and Economics of Public Goods
**Description [Round 1]**: This cluster focuses on the economic and political dynamics associated with the provision of public goods. Public goods are typically defined by characteristics such as non-excludability and non-rivalry in consumption. Essential themes within this cluster include the allocation and efficient provision of public goods, the role of taxation and fiscal policies, and the impact of public and private sector contributions. The cluster also delves into the comparative efficacy of centralized versus decentralized systems in public goods provision, the influence of electoral incentives, and the effect of ethnic diversity on public goods accessibility. Other discussed topics involve equilibrium, efficiency, and government expenditure, and the role of bargaining, voter preferences, and federalism in shaping public goods distribution.

Additionally, social and economic factors like income distribution, taxation, and the interplay between private and public goods further enrich the discourse.
**Description [Round 2]**: This cluster explores the complex dynamics surrounding the provision and distribution of public goods within various economic systems. It encompasses theoretical approaches to the provision of public goods, debates between centralized versus decentralized provision, and the influence of ethnic and socio-economic diversity on public goods. Key themes include the role of taxation, government expenditure, and economic efficiency, as well as issues of income distribution, jurisdiction, decentralization, and the political economy. The cluster also examines mathematical and theoretical frameworks such as equilibrium analysis, Pareto efficiency, and Samuelson's model, in addition to real-world concerns like tax competition, voter preferences, and government incentives.
**Top 100 grams (The number in parentheses is the value of relative frequency.)**:
public good(0.64), good(0.55), public(0.43), provision(0.32), good provision(0.19), local public(0.17), goods(0.15), public goods(0.14), provision public(0.13), tax(0.10), local(0.10), equilibrium(0.09), optimal(0.09), economy(0.08), taxation(0.08), preference(0.07), fiscal(0.06), efficient(0.06), spending(0.06), allocation(0.06), taxis(0.06), government(0.06), efficiency(0.06), expenditure(0.05), income(0.05), property(0.05), welfare(0.05), cost(0.05), level public(0.05), jurisdiction(0.05), voter(0.05), private good(0.05), externality(0.05), local government(0.05), rule(0.04), incentive(0.04), elsevi right(0.04), finance(0.04), provide public(0.04), private(0.04), price(0.04), excludable(0.04), good economy(0.04), provision local(0.04), decentralization(0.04), elsevi(0.04), reserve(0.04), right reserve(0.04), pareto(0.04), marginal(0.03), good public(0.03), valuation(0.03), spillover(0.03), political economy(0.03), excludable public(0.03), budget(0.03), agent(0.03), bargaining(0.03), willingness pay(0.03), elsevier science(0.03), tax rate(0.03), ethnic(0.03), voting(0.03), competition(0.03), supply(0.03), municipality(0.03), federalism(0.03), public service(0.03), income tax(0.03), goods provision(0.03), preference public(0.03), utility(0.03), distribution(0.03), electoral(0.03), election(0.03), public spending(0.03), vote(0.03), samuelson(0.03), redistribution(0.02), decentralized(0.02), district(0.02), proof(0.02), ethnic diversity(0.02), tiebout(0.02), estimate(0.02), household(0.02), pure(0.02), cost public(0.02), satisfy(0.02), economy public(0.02), tax competition(0.02), median(0.02), theorem(0.02), mechanism design(0.02), good provide(0.02), public expenditure(0.02), strategy proof(0.02), village(0.02), taxation public(0.02), cost sharing(0.02),
**Titles of the top 10 most central articles**:
1. Public Goods And Ethnic Divisions
2. Private Demands For Public Goods
3. Lindahls Solution And Core Of An Economy With Public Goods
4. Centralized Versus Decentralized Provision Of Local Public Goods: A Political Economy Approach
5. The Provision Of Public Goods Under Alternative Electoral Incentives
6. Pigou, Taxation And Public Goods
7. Ethnic Diversity, Social Sanctions, And Public Goods In Kenya
8. Does Voluntary Participation Undermine The Coase Theorem?
9. It's Not Just Welfare: Racial Inequality And The Local Provision Of Public Goods In The United States
10. Public Goods And The Distribution Of Income

***Cluster 10***
**Title [Round 1]**: Corporate Philanthropy and Its Strategic Implications

**Title [Round 2]**: Corporate Philanthropy and Social Responsibility Dynamics
**Description [Round 1]**: This cluster focuses on the intersection of corporate philanthropy and strategic business practices. It examines how corporate social responsibility (CSR) initiatives and philanthropic activities can influence a company's financial performance, reputation, and consumer perceptions. Key areas of interest include strategic philanthropy, cause-related marketing, and corporate donations. The cluster also explores the implications of corporate giving on stakeholder relations, marketing strategies, and corporate governance. Evidence from various industries and regions, particularly in China, is discussed to assess the impact of these initiatives on the firm's competitive advantage and the associated ethical considerations.
**Description [Round 2]**: This cluster explores the intricate relationship between corporate philanthropy and corporate social responsibility (CSR), emphasizing the influence on firms, consumers, and stakeholders. It covers various aspects including strategic philanthropy, cause-related marketing, and the impact of corporate giving on business reputation and financial performance. The integration of CSR activities into business strategies is analyzed, with a focus on consumer perceptions, shareholder responses, and corporate governance. The geographical context, particularly insights from Chinese firms, is also considered, alongside the methodological approaches employed in philanthropy-related research.
**Top 100 grams (The number in parentheses is the value of relative frequency.)**:
corporate(0.65), philanthropy(0.42), firm(0.35), corporate philanthropy(0.34), corporate social(0.33), social responsibility(0.33), responsibility(0.32), company(0.25), philanthropic(0.23), csr(0.23), responsibility csr(0.18), business(0.15), consumer(0.13), giving(0.11), list(0.10), corporation(0.10), marketing(0.09), donation(0.09), charitable(0.09), chinese(0.09), corporate philanthropic(0.08), corporate giving(0.08), strategic(0.08), stakeholder(0.08), china(0.07), philanthropy corporate(0.07), industry(0.07), cause relate(0.07), originality(0.07), brand(0.07), moderate(0.06), relate marketing(0.06), design methodology(0.06), csr activity(0.06), originality value(0.06), methodology approach(0.06), cause(0.06), manager(0.05), ownership(0.05), corporate charitable(0.05), philanthropic activity(0.05), performance(0.05), relationship corporate(0.05), reputation(0.05), philanthropic giving(0.04), financial performance(0.04), methodology(0.04), managerial(0.04), responsibility corporate(0.04), stock(0.04), own(0.04), practical implication(0.04), effect corporate(0.04), list firm(0.04), chinese list(0.04), sponsorship(0.04), corporate donation(0.04), findings(0.04), customer(0.04), product(0.04), charitable donation(0.04), ethical(0.04), crm(0.04), shareholder(0.04), attribution(0.03), positively(0.03), motive(0.03), purchase(0.03), research limitation(0.03), purpose paper(0.03), fit(0.03), corporate foundation(0.03), ceo(0.03), moderate effect(0.03), marketing crm(0.03), influence corporate(0.03), sponsor(0.03), responsible(0.03), practical(0.03), state own(0.03), list company(0.03), corporate governance(0.03), purchase intention(0.03), limitation implication(0.03), giving corporate(0.03), csr corporate(0.03), donate(0.03), consumer perception(0.03), evidence china(0.03), political connection(0.03), impact corporate(0.03), influence consumer(0.03), consumer attitude(0.03), socially responsible(0.03), advertising(0.03), strategic philanthropy(0.03), firm corporate(0.03), multinational(0.02), charitable giving(0.02), private firm(0.02),
**Titles of the top 10 most central articles**:
1. The Competitive Advantage Of Corporate Philanthropy
2. Research On Corporate Philanthropy: A Review And Assessment
3. Cause-Related Marketing - A Coalignment Of Marketing Strategy And Corporate Philanthropy

4. Corporate Philanthropy And Corporate Financial Performance: The Roles Of Stakeholder Response And Political Access
5. Battling The Devolution In The Research On Corporate Philanthropy
6. Is Doing Good Good For You? How Corporate Charitable Contributions Enhance Revenue Growth
7. The Effect Of Corporate Social Responsibility On Customer Donations To Corporate-Supported Nonprofits
8. Corporate Reputation And Philanthropy: An Empirical Analysis
9. Building Corporate Associations: Consumer Attributions For Corporate Socially Responsible Programs
10. The New Corporate Philanthropy

***Cluster 11***
**Title [Round 1]**: Role of Self-Help and Mutual Aid Groups in Health and Empowerment
**Title [Round 2]**: The Role of Self-Help and Mutual Aid Groups in Mental Health and Substance Abuse Recovery
**Description [Round 1]**: This cluster explores the significance and impact of self-help and mutual aid groups, with a focus on their role in supporting mental health, substance abuse recovery, and patient empowerment. It delves into the psychological and social processes within these groups, their effectiveness, and their contribution to the well-being of individuals facing addiction, chronic illnesses, and various mental health issues. Additionally, the cluster examines the existence and function of these groups across different cultural contexts, particularly in rural India, and their influence as development intermediaries in South Asia and Africa. It also considers the support mechanisms within these groups, including peer support and organizational exchange with mental health and substance use agencies.
**Description [Round 2]**: This cluster explores the impact and dynamics of self-help and mutual aid groups on personal recovery and empowerment within mental health and substance use contexts. Keywords such as 'self-help', 'help group', and 'mutual aid' dominate this cluster, indicating a focus on voluntary support systems leveraged by individuals with shared experiences. The analysis covers the psychological benefits of these groups, their contributions to mental health well-being, and their structure as community-driven initiatives. Sample titles suggest an examination of experiential knowledge shared within these groups and their integration with formal mental health and substance abuse agencies. The geographic and demographic focus likely includes broad regions such as rural India, highlighting diversity in group implementations and outcomes. Research methodologies such as cross-sectional analyses and case studies characterize the academic approach to understanding these grassroots movements.
**Top 100 grams (The number in parentheses is the value of relative frequency.)**:
self help(0.73), help group(0.56), self(0.56), help(0.53), group(0.37), mutual(0.22), groups(0.20), mutual aid(0.19), help groups(0.18), patient(0.18), health(0.18), aid(0.16), member(0.15), treatment(0.14), shg(0.13), recovery(0.13), shgs(0.12), mental(0.11), india(0.10), woman(0.10), substance(0.10), care(0.09), mental health(0.09), empowerment(0.09), attend(0.09), alcohol(0.09), professional(0.08), group shgs(0.08), support group(0.08), anonymous(0.08), meeting(0.08), disorder(0.08), attendance(0.07), drug(0.07), addiction(0.07), alcoholics(0.07), illness(0.07), substance use(0.07), alcoholics anonymous(0.07), step(0.07), aid group(0.06), help organization(0.06), cancer(0.06), intervention(0.06), disease(0.06), month(0.05), cope(0.05), use disorder(0.05), conclusion(0.05), health care(0.05), social support(0.05), group member(0.05),

abstinence(0.05), involvement(0.05), background(0.05), abuse(0.05), program(0.05), rehabilitation(0.05), women(0.05), referral(0.04), woman empowerment(0.04), group shg(0.04), participation self(0.04), shg member(0.04), group participation(0.04), woman self(0.04), clinical(0.04), psychiatric(0.04), chronic(0.04), group self(0.04), step group(0.04), method(0.04), peer(0.04), health service(0.04), mental illness(0.04), conclusions(0.04), peer support(0.04), substance abuse(0.04), rural(0.03), empower(0.03), support self(0.03), microfinance(0.03), diagnosis(0.03), help mutual(0.03), mutual help(0.03), baseline(0.03), outpatient(0.03), therapy(0.03), member self(0.03), participation(0.03), step self(0.03), group patient(0.03), therapeutic(0.03), medical(0.03), cross sectional(0.03), sectional(0.03), rural india(0.03), alcoholic(0.03), questionnaire(0.03), membership(0.03),
**Titles of the top 10 most central articles**:
1. The Contribution Of Self-Help/Mutual Aid Groups To Mental Well-Being
2. Experiential Knowledge - New Concept For Analysis Of Self-Help Groups
3. Mutual Help Groups For Mental Health Problems: A Review Of Effectiveness Studies
4. Self-Help Organizations For Alcohol And Drug Problems: Toward Evidence-Based Practice And Policy
5. Self-Help Groups - Types And Psychological Processes
6. Self-Help Groups And Mental Health/Substance Use Agencies: The Benefits Of Organizational Exchange
7. Solidarity Not Charity Mutual Aid For Mobilization And Survival
8. Political Ideology, Helping Mechanisms And Empowerment Of Mental Health Self-Help/Mutual Aid Groups
9. Typologies Of Mutual Aid In Climate Resilience: Variation In Reciprocity, Solidarity, Self-Determination, And Resistance
10. Delivering Development? Evidence On Self-Help Groups As Developme

***Cluster 12***
**Title [Round 1]**: Prosocial Behavior and Motivation in the Workplace
**Title [Round 2]**: Prosocial Behaviors and Motivations in the Workplace
**Description [Round 1]**: This cluster focuses on the study of employee behavior, particularly how intrinsic and prosocial motivations impact job performance, relationships, and organizational dynamics. It explores how prosocial motivations can mediate and moderate employee behaviors, such as helping and citizenship actions, within workplace structures. The cluster examines the role of leadership, supervisor influence, and trust in fostering a positive work environment and the subsequent effects on organizational citizenship behaviors (OCB). Using methodologies like structural equation modeling, the research looks into both direct and indirect effects of motivation on performance outcomes and evaluates theoretical and practical implications for workplaces across various industries.
**Description [Round 2]**: This cluster focuses on the intricacies of employee behavior within organizational settings, particularly highlighting the role of prosocial motivation. It examines the relationship dynamics involving supervisors, managers, and coworkers, and addresses the mediating effects of prosocial behaviors on overall performance and organizational citizenship. Using methodologies such as structural equation modeling, studies in this group shed light on the influence of motivational synergies on employee persistence, productivity, and job satisfaction. Moreover, this cluster investigates the implications of social exchange and the impact of perceived leadership on encouraging positive workplace behaviors. Key discussions emphasize the mediating roles and boundary conditions that affect prosocial employee conduct in diverse industries, including service and hospitality. Empirical findings are frequently supported by methodologies like questionnaires and data collection, contributing valuable insights into practical and theoretical applications.

**Top 100 grams (The number in parentheses is the value of relative frequency.)**: employee(0.42), behavior(0.19), relationship(0.18), mediate(0.17), help behavior(0.17), prosocial(0.16), moderate(0.15), motivation(0.14), organizational(0.14), collect(0.14), job(0.13), prosocial motivation(0.13), supervisor(0.12), performance(0.12), positively(0.12), citizenship(0.11), structural equation(0.10), citizenship behavior(0.10), workplace(0.10), datum collect(0.10), practical implication(0.10), equation(0.10), work(0.09), positive(0.09), organizational citizenship(0.09), test(0.09), mediate relationship(0.09), leadership(0.09), practical(0.08), mediating(0.08), perceive(0.08), implication(0.08), antecedent(0.08), help(0.07), structural(0.07), customer(0.07), leader(0.07), mediation(0.07), pro(0.06), manager(0.06), team(0.06), originality(0.06), design methodology(0.06), propose(0.06), helping(0.06), positively relate(0.06), modeling(0.06), hypothesis(0.06), effect(0.06), moderated(0.06), equation modeling(0.06), ocb(0.06), originality value(0.06), behaviour(0.06), methodology approach(0.06), coworker(0.06), indirect(0.05), theory(0.05), theoretical practical(0.05), exchange(0.05), perception(0.05), task(0.05), indirect effect(0.05), social exchange(0.05), moderate effect(0.05), employee work(0.05), benevolence(0.05), moderated mediation(0.05), positive relationship(0.05), moderate relationship(0.05), moderate role(0.05), subordinate(0.05), mediation model(0.05), mediating role(0.04), employee prosocial(0.04), proactive(0.04), employee help(0.04), hotel(0.04), questionnaire(0.04), conservation resource(0.04), intention(0.04), extra role(0.04), behavior ocb(0.04), extra(0.04), rule breaking(0.04), breaking(0.04), psychological(0.04), mediator(0.04), managerial(0.04), altruistic value(0.04), negative(0.04), influence(0.04), affective(0.04), mediate role(0.04), behavior employee(0.04), resource theory(0.04), industry(0.04), rule break(0.04), help behaviour(0.04), boundary condition(0.04),

**Titles of the top 10 most central articles**:

1. Does Intrinsic Motivation Fuel The Prosocial Fire? Motivational Synergy In Predicting Persistence, Performance, And Productivity
2. Pro-Social Organizational Behaviors
3. The Necessity Of Others Is The Mother Of Invention: Intrinsic And Prosocial Motivations, Perspective Taking, And Creativity
4. The Synergistic Impact Of Motivations On Sustained Pro-Environmental Consumer Behaviors: An Empirical Evidence For Single-Use Plastic Products
5. Understanding Prosocial Behavior, Sales Performance, And Turnover - A Group-Level Analysis In A Service Context
6. The Influence Of Vietnamese Consumers' Altruistic Values On Their Purchase Of Energy Efficient Appliances
7. Feeling Bad And Doing Good: The Effect Of Customer Mistreatment On Service Employee's Daily Display Of Helping Behaviors
8. Nursing Students' Prosocial Motivation: Does It Predict Professional Commitment And Involvement In The Job?
9. Mission Possible? The Performance Of Prosocially Motivated Employees Depends On Manager Trustworthiness
10. Feeling Good, Doing Good, And Getting Ahead: A Meta-Analytic Investigation Of The Outcomes Of Prosocial Motivation At Work

***Cluster 13***
**Title [Round 1]**: Islamic Waqf and Charity Practices
**Title [Round 2]**: Exploring the Role of Waqf in Islamic Philanthropy and Socio-Economic Development
**Description [Round 1]**: This cluster primarily focuses on the Islamic concept of Waqf, which refers to a religious endowment specifically designated for charitable purposes. It highlights the significance of charity and zakat (almsgiving) in Islam, illustrating how these practices are institutionalized within Muslim communities across different centuries. Key topics include the development and management of Waqf institutions in regions such as Malaysia and Indonesia, the legal frameworks surrounding Waqf properties, and the socio-economic impacts of Islamic philanthropy. This cluster also explores historical contexts, such as Waqf's role in the Ottoman Empire and colonial South Asia, and its contemporary applications in poverty alleviation and Islamic finance.
**Description [Round 2]**: This cluster focuses on the institution of waqf, a form of Islamic endowment, and its significance in charity and socio-economic development. The analysis covers various historical and contemporary contexts, including its implementation in countries like Malaysia and Indonesia, and its broader implications for poverty alleviation within the Muslim community. The research highlights the operational methodologies of waqf, its integration with modern Islamic finance, and the socio-religious motivations behind its establishment and management. Additionally, the historical evolution of waqf, particularly during the Ottoman Empire and in colonial contexts, is examined, offering insights into the ongoing relevance and challenges faced by waqf institutions.
**Top 100 grams (The number in parentheses is the value of relative frequency.)**:
waqf(0.30), islamic(0.28), charity(0.20), zakat(0.17), institution(0.13), muslim(0.13), religious(0.13), century(0.12), malaysia(0.12), originality(0.09), design methodology(0.09), poor(0.09), islam(0.08), endowment(0.08), originality value(0.08), poverty(0.08), methodology approach(0.08), modern(0.07), methodology(0.07), waqf institution(0.07), finance(0.07), cash waqf(0.07), cash(0.07), findings(0.06), muslims(0.06), research limitation(0.06), philanthropy(0.06), indonesia(0.06), limitation implication(0.05), waqf islamic(0.05), collection(0.05), paper aim(0.05), fund(0.05), malaysian(0.04), ottoman(0.04), financing(0.04), late(0.04), purpose(0.04), zakat institution(0.04), law(0.04), purpose paper(0.04), colonial(0.04), relief(0.04), charitable(0.04), practical implication(0.04), land(0.04), wealth(0.03), early modern(0.03), limitation(0.03), contemporary(0.03), legal(0.03), development waqf(0.03), islamic finance(0.03), poverty alleviation(0.03), shari'ah(0.03), waqf land(0.03), waqf fund(0.03), socio economic(0.03), alleviation(0.03), empire(0.03), alm(0.03), property(0.03), britain(0.03), islamic endowment(0.03), waqf model(0.03), waqfs(0.03), waqf management(0.03), islamic religious(0.03), intellectual capital(0.03), establishment(0.03), intellectual(0.03), indonesian(0.03), islamic financial(0.03), gift(0.03), awqaf(0.03), islamic economic(0.03), asset(0.03), religion(0.02), islamic social(0.02), history(0.02), london(0.02), spiritual(0.02), pious(0.02), distribution(0.02), nineteenth(0.02), book(0.02), shariah(0.02), waqf asset(0.02), humanitarianism(0.02), nineteenth century(0.02), patronage(0.02), proper(0.02), social finance(0.02), muslim community(0.02), war(0.02), financial(0.02), aid(0.02), zakat islamic(0.02), humanitarian(0.02), beneficiary(0.02),
**Titles of the top 10 most central articles**:
1. Critical Assessment Of Islamic Endowment Funds (Waqf) Literature: Lesson For Government And Future Directions

2. Gifting And Receiving: Anglo-Indian Charity And Its Beneficiaries In Madras
3. Logic Of Charity - Poor Relief In Preindustrial Europe
4. Philanthropy And The Social History Paradigm
5. Property, Authority And Personal Law: Waqf In Colonial South Asia
6. Charity And Philanthropy In South Asia: An Introduction
7. Financial Worship: The Quranic Injunction To Almsgiving
8. From Tribute To Philanthropy - The Politics Of Gift Giving In A Western Indian City
9. 'you Can Give Even If You Only Have Ten Rupees!': Muslim Charity In A Colombo Housing Scheme
10. The Impulse Of Philanthropy

***Cluster 14***
**Title [Round 1]**: Intergenerational Altruism and Economic Behavior
**Title [Round 2]**: Altruism and Intergenerational Transfers in Economic Models
**Description [Round 1]**: This cluster explores the concept of altruism with a particular focus on intergenerational dynamics and economic implications. Key themes include the equilibrium of altruistic behaviors in family and economic settings, the transfer of resources, and parental preferences towards their children. Various economic models and constructs, such as the overlapping generation model, are discussed to examine how altruistic behaviors can influence utility, welfare, and social security. The cluster also touches on the role of altruistic motives in consumption, savings, and economic growth, providing insights into the theoretical underpinnings of altruistic preferences and their practical impact on family and market economics.
**Description [Round 2]**: This cluster explores the concept of altruism and its impact within economic frameworks, particularly focusing on intergenerational transfers, family dynamics, and economic equilibria. Key topics include the study of altruistic behaviors, how such behaviors contribute to household and familial utility, and the broader implications on economic systems. The cluster examines models that incorporate altruistic preferences, intergenerational dependencies, and the equilibrium outcomes of altruistic interactions. Additionally, the discussions extend to the mechanisms of transfer and bequest, the role of parental preferences, and the implications on welfare and policy. The interplay between altruism and economic growth, as well as the related effects on consumption, saving, and income distribution, also form a significant part of this body of work. Further analysis involves exploring concepts like Nash equilibrium, overlapping generation models, and externalities in the context of altruistic actions within economic systems.
**Top 100 grams (The number in parentheses is the value of relative frequency.)**: altruism(0.50), altruistic(0.28), equilibrium(0.20), transfer(0.18), parent(0.14), preference(0.14), intergenerational(0.14), child(0.12), generation(0.11), family(0.10), utility(0.10), household(0.10), bequest(0.09), optimal(0.09), agent(0.08), consumption(0.08), parental(0.07), model(0.07), altruistic preference(0.07), right reserve(0.06), supply(0.06), elsevi right(0.06), reserve(0.06), income(0.06), endogenous(0.06), welfare(0.06), intergenerational transfer(0.06), willingness pay(0.05), overlapping generation(0.05), degree altruism(0.05), overlapping(0.05), valuation(0.05), externality(0.05), perfect(0.05), degree(0.05), insurance(0.05), intergenerational altruism(0.05), elsevi(0.05), price(0.05), supply chain(0.04), parent child(0.04), motive(0.04), long run(0.04), depend(0.04), generation model(0.04), parental altruism(0.04), overlap generation(0.04), allocation(0.04), dynastic(0.04), pareto(0.04), nash(0.04), chain(0.04), fertility(0.04), social security(0.04), estimate(0.04), existence(0.04), saving(0.04), wtp(0.03),

pay(0.03), overlap(0.03), private transfer(0.03), remittance(0.03), tax(0.03), willingness(0.03), stationary(0.03), game(0.03), inheritance(0.03), utility function(0.03), pure(0.03), economic growth(0.03), run(0.03), manufacturer(0.03), altruistic parent(0.03), equilibria(0.03), elsevier(0.03), transfer altruism(0.03), nash equilibrium(0.03), pay wtp(0.03), paper study(0.03), standard(0.03), altruism family(0.03), weight(0.03), taxis(0.03), carbon(0.03), capital accumulation(0.03), contingent valuation(0.03), inter vivos(0.03), numerical(0.03), vivos(0.03), reduction(0.03), paternalistic(0.03), elsevier science(0.03), accumulation(0.03), marginal(0.02), bequ(0.02), fiscal policy(0.02), theorem(0.02), markov(0.02), retailer(0.02), transfer child(0.02),
**Titles of the top 10 most central articles**:
1. Altruism In Preventive Health Behavior: At-Scale Evidence From The Hiv/Aids Pandemic
2. Altruism, Egoism, And Genetic Fitness - Economics And Sociobiology
3. Altruism And Time Consistency - The Economics Of Fait Accompli
4. Parental Altruism And Inter Vivos Transfers: Theory And Evidence
5. Altruism In The Family And Selfishness In The Market Place
6. Altruism In Networks
7. Altruism Within The Family Reconsidered - Do Nice Guys Finish Last
8. Public Goods In Networks
9. Is Altruism Evolutionarily Stable?
10. Non-Paternalistic Altruism And Welfare Economics

***Cluster 15***
**Title [Round 1]**: The Dynamics of the Commons and Public Goods
**Title [Round 2]**: Dynamics of Commons and Public Goods Management
**Description [Round 1]**: This cluster explores the concept of the commons, the tragedy often associated with them, and their governance as public goods. It delves into the mechanisms and theories of property rights, resource management, and environmental conservation. Key discussions revolve around the challenges and methodologies in managing common resources like land, water, and fisheries, addressing issues of privatization, ecological sustainability, and institutional governance. Influential scholars such as Elinor Ostrom and Garrett Hardin feature prominently in the discourse, contributing to an understanding of collective action and socio-ecological systems.
**Description [Round 2]**: This cluster delves into the complexities surrounding the concept of the commons, highlighting themes such as public goods, property rights, and resource management. It explores the challenges and dynamics inherent in the management and governance of shared resources, often referred to as 'the tragedy of the commons', a term popularized by Hardin. The cluster examines the interplay between private and public ownership, common pool resources, and governance mechanisms necessary to avoid depletion. Influence from thought leaders like Elinor Ostrom is notable, providing insights on sustainable and resilient governance models. The interplay between ecological, economic, and social factors in resource management is a significant area of focus, with implications for global public goods and environmental conservation strategies.
**Top 100 grams (The number in parentheses is the value of relative frequency.)**:
common(0.43), commons(0.27), tragedy(0.21), public good(0.19), property(0.17), tragedy common(0.16), good(0.14), resource(0.14), common pool(0.13), governance(0.12), pool resource(0.12), pool(0.12), land(0.10), natural(0.09), global(0.09), property right(0.09), management(0.09), natural resource(0.08), global public(0.08), environmental(0.08), private(0.08), tragedy commons(0.08), ecological(0.07), common property(0.07), fishery(0.06),

ostrom(0.06), conservation(0.06), access(0.06), collective(0.06), high education(0.06), agricultural(0.06), govern(0.06), resource management(0.05), agriculture(0.05), collective action(0.05), hardin(0.05), social ecological(0.05), ecosystem(0.05), forest(0.05), water(0.04), institution(0.04), biodiversity(0.04), production(0.04), farmer(0.04), system(0.04), ecological system(0.04), privatization(0.04), institutional(0.04), climate change(0.04), ownership(0.04), elinor ostrom(0.04), elinor(0.04), agreement(0.04), regime(0.04), goods(0.04), sustainable(0.04), valuation(0.03), protect(0.03), externality(0.03), arrangement(0.03), private property(0.03), open access(0.03), public goods(0.03), climate(0.03), protection(0.03), degradation(0.03), ecosystem service(0.03), economic(0.03), fishing(0.03), education public(0.03), common resource(0.03), contingent valuation(0.03), manage(0.03), agricultural policy(0.03), management common(0.03), european union(0.03), willingness pay(0.03), farming(0.03), common good(0.03), public private(0.03), regulation(0.03), solution(0.02), mitigation(0.02), valuation method(0.02), open(0.02), marine(0.02), agri(0.02), resource system(0.02), communal(0.02), govern common(0.02), stock(0.02), farm(0.02), free(0.02), policy(0.02), political economy(0.02), good public(0.02), landscape(0.02), payment(0.02), law(0.02), area(0.02),
**Titles of the top 10 most central articles**:
1. The Tragedy Of The Commons - 22 Years Later
2. Governing As Commons Or As Global Public Goods: Two Tales Of Power
3. Global Public Goods And Democracy In International Legal Scholarship
4. Public Goods, Global Public Goods And The Common Good
5. Polycentric And Resilient Perspectives For Governing The Commons: Strategic And Law And Economics Insights For Sustainable Development
6. The Comedy Of The Commons - Custom, Commerce, And Inherently Public Property
7. Ostrom, Hardin And The Commons: A Critical Appreciation And A Revisionist View
8. Retrospectives Tragedy Of The Commons After 50 Years
9. A Constitutional Theory Of Public Goods
10. Ostrom's Law: Property Rights In The Commons

***Cluster 16***
**Title [Round 1]**: Social Media Utilization by Nonprofit Organizations
**Title [Round 2]**: Role of Social Media in Nonprofit Organization Communication
**Description [Round 1]**: This cluster explores the integration and application of social media platforms such as Facebook and Twitter by nonprofit and nongovernmental organizations. It delves into how these mediums serve as effective communication channels to engage stakeholders, enhance public relations, and facilitate fundraising efforts. The discussion includes an analysis of online content strategies, social marketing techniques, and the use of technology to bolster organizational advocacy campaigns. The efficacy of websites as a tool for stakeholder engagement and the role of information communication technology in fostering dialogic interactions are also examined.
**Description [Round 2]**: This cluster explores how nonprofit and non-governmental organizations leverage social media platforms like Facebook and Twitter for communication and stakeholder engagement. It delves into the role of these platforms in fundraising, advocacy, and the promotion of organizational causes. The cluster examines the effectiveness of various social marketing strategies, digital communication technologies, and the impact of online presence on public relations and audience interaction in the nonprofit sector.
**Top 100 grams (The number in parentheses is the value of relative frequency.)**:

medium(0.33), communication(0.31), social medium(0.30), nonprofit(0.17), organization(0.15), information(0.14), online(0.14), nonprofit organization(0.13), facebook(0.13), technology(0.12), public relation(0.12), media(0.12), twitter(0.12), marketing(0.12), social media(0.11), content(0.11), message(0.10), use(0.10), stakeholder(0.09), social marketing(0.08), digital(0.08), internet(0.08), content analysis(0.08), web(0.07), website(0.07), fundraising(0.07), use social(0.07), social network(0.06), organizations(0.06), non profit(0.06), user(0.06), platform(0.06), site(0.06), ngos(0.06), non governmental(0.06), network(0.05), communicate(0.05), advocacy(0.05), profit organization(0.05), campaign(0.05), audience(0.05), dialogic(0.05), tool(0.05), tweet(0.05), adoption(0.05), governmental(0.04), organization social(0.04), npos(0.04), organization use(0.04), organization public(0.04), information technology(0.04), communicative(0.04), governmental organization(0.04), medium use(0.04), npo(0.04), ict(0.04), marketer(0.04), ngo(0.04), stakeholder engagement(0.04), practitioner(0.03), medium platform(0.03), communication technology(0.03), information communication(0.03), nongovernmental(0.03), network analysis(0.03), technological(0.03), utilize(0.03), communication strategy(0.03), facebook twitter(0.03), follower(0.03), organizational(0.03), relation(0.03), nonprofit organizations(0.03), communication social(0.03), awareness(0.03), inc.(0.03), relationship management(0.03), interactivity(0.03), network site(0.03), dialogic communication(0.03), web site(0.03), social marketer(0.03), page(0.03), use twitter(0.03), strategic communication(0.03), organization npos(0.03), channel(0.03), social networking(0.02), nongovernmental organization(0.02), usage(0.02), post(0.02), profit sector(0.02), profit(0.02), engagement(0.02), networking(0.02), public relationship(0.02), organization ngos(0.02), way communication(0.02), medium twitter(0.02), dialogue(0.02),
**Titles of the top 10 most central articles**:
1. The Who, Where, And When Of Social Marketing
2. Information, Community, And Action: How Nonprofit Organizations Use Social Media
3. Engaging Stakeholders Through Social Networking: How Nonprofit Organizations Are Using Facebook
4. Tweets For Tots: Using Twitter To Promote A Charity And Its Supporters
5. Are The Non-Governmental Organizations' Web Sites Effective?
6. Competing Voices: Marketing And Counter-Marketing Alcohol On Twitter
7. Global Cause Awareness: Tracking Awareness Through Electronic Word Of Mouth
8. Engaging Stakeholders Through Twitter: How Nonprofit Organizations Are Getting More Out Of 140 Characters Or Less
9. On The Effectiveness Of Social Marketing-What Do We Really Know?
10. Unpacking The Drivers Of Stakeholder Engagement In Sustainable Water Management: Ngos And The Use Of Facebook

***Cluster 17***
**Title [Round 1]**: Role of Faith-Based Organizations in Voluntary Social Services
**Title [Round 2]**: Role of Faith-Based Voluntary Organizations in Social Services
**Description [Round 1]**: This cluster examines the interplay between faith-based organizations and the voluntary sector in providing social services and alleviating food insecurity. It focuses on the role of religious and voluntary groups, primarily in Canada, the UK, and Australia, as they navigate charity-driven efforts such as food banks and pantries. These organizations contribute significantly to addressing welfare needs in various geographical contexts, often operating under the framework of faith, voluntarism, and community engagement. Research explores their impact on the provision of social services, partnership with secular organizations, and the challenges posed by contemporary societal structures like neoliberalism and austerity.

**Description [Round 2]**: This cluster explores the intersection of faith, voluntarism, and social services within various communities. Emphasizing the role of faith-based organizations, it delves into how religious motives fuel voluntary efforts, particularly in addressing food insecurity through food banks and charities. The cluster analyzes the structural components and motivations behind these initiatives, examining the nuanced ways in which they contribute to welfare and social services. Geographic considerations, particularly in regions like Canada and England, are discussed, as well as the interplay between secular and religious efforts in the voluntary sector. Consideration is given to how these organizations operate within the social safety net, addressing emergencies, homeless services, and partnering with government bodies amidst social policy reforms and the challenges posed by neoliberal economic frameworks.

**Top 100 grams (The number in parentheses is the value of relative frequency.)**:
faith(0.18), faith base(0.16), voluntary sector(0.16), food(0.15), service(0.15), voluntary(0.13), base organization(0.10), food bank(0.10), welfare(0.10), social service(0.09), religious(0.09), voluntarism(0.09), bank(0.09), charity(0.09), insecurity(0.08), food insecurity(0.08), charitable(0.08), assistance(0.08), geography(0.07), canada(0.07), religion(0.07), sector(0.06), provision(0.06), church(0.06), secular(0.06), provider(0.06), poverty(0.06), agency(0.06), charitable choice(0.06), england(0.05), charitable food(0.05), delivery(0.05), food assistance(0.05), client(0.05), christian(0.05), service provision(0.05), hunger(0.05), organisation(0.05), food security(0.05), community(0.04), food charity(0.04), restructuring(0.04), fbo(0.04), funding(0.04), charity food(0.04), emergency(0.04), congregation(0.04), rural(0.04), ontario(0.04), role voluntary(0.04), relief(0.04), initiative(0.03), catholic(0.03), welfare reform(0.03), security(0.03), city(0.03), welfare state(0.03), criminal justice(0.03), partnership(0.03), emergency food(0.03), homeless(0.03), discourse(0.03), penal(0.03), australia(0.03), place(0.03), meet(0.03), poor(0.03), criminal(0.03), base initiative(0.03), bank food(0.03), jewish(0.03), food aid(0.03), pantry(0.03), program(0.03), urban(0.03), rural community(0.03), meal(0.03), serve(0.03), penal voluntary(0.03), service provider(0.03), food pantry(0.03), service delivery(0.03), neoliberal(0.03), base organizations(0.02), voluntary organization(0.02), operation(0.02), food system(0.02), austerity(0.02), government(0.02), geographical(0.02), reform(0.02), social work(0.02), centre(0.02), social policy(0.02), services(0.02), religious organization(0.02), volunteer(0.02), homelessness(0.02), experience food(0.02), organization fbo(0.02),

**Titles of the top 10 most central articles**:
1. Out Of The Shadows: Exploring Contemporary Geographies Of Voluntarism
2. Typology Of Religious Characteristics Of Social Service And Educational Organizations And Programs
3. Religious Faith, Effort And Enthusiasm: Motivations To Volunteer In Response To Holiday Hunger
4. Victim Support, The State, And Fellow Human Beings
5. Unintended Consequence Of The Faith-Based Initiative: Organizational Practices And Religious Identity Within Faith-Based Human Service Organizations
6. Charitable Choice And Faith-Based Welfare: A Call For Social Work
7. Towards A Relational View Of The Shadow State
8. The Relational Geographies Of The Voluntary Sector: Disentangling The Ballast Of Strangers

9. Countermovement, Neoliberal Platoon, Or Re-Gifting Depot? Understanding Decommodification In Us Food Banks
10. Defining Faith-Based Organizations And Understanding Them Through Research

***Cluster 18***
**Title [Round 1]**: Role of Non-Governmental Organizations in Global Health Care and HIV Prevention
**Title [Round 2]**: Role and Impact of Non-Governmental Organizations in Global Health
**Description [Round 1]**: This cluster explores the involvement of non-governmental organizations (NGOs) and community-based organizations (CBOs) in delivering health services and implementing HIV prevention programs globally, especially in low- and middle-income countries. It discusses the collaborative efforts between NGOs, government bodies, and international partners in addressing healthcare disparities, providing medical care, and enhancing health systems. The literature examines the challenges and strategies in strengthening NGO participation in the healthcare sector and evaluates their impact through evidence-based interventions and volunteerism. The cluster provides insights into the effectiveness of NGO-driven initiatives, the capacity building of these organizations, and their critical role in improving global health outcomes.
**Description [Round 2]**: This cluster examines the influence and contributions of non-governmental organizations (NGOs) within the health sector on a global scale. It highlights the partnerships between public entities and NGOs, with focus on their roles in promoting health practices, providing medical care, and implementing health programs in various settings, particularly in low- and middle-income countries. The cluster explores how these organizations participate in HIV prevention, support healthcare delivery, and address health disparities. By discussing strategies for improvement and collaboration, this collection of works emphasizes the potential of NGOs to enhance public health systems, strengthen community-based health initiatives, and contribute to surgical volunteerism and medical care in regions like Africa and Ukraine. Challenges and successes of NGOs in global health and their partnership with governmental bodies are also reflected through a series of case studies and program analyses.
**Top 100 grams (The number in parentheses is the value of relative frequency.)**:
health(0.46), community base(0.28), care(0.24), base organization(0.23), hiv(0.23), community(0.21), non governmental(0.19), governmental(0.19), service(0.17), cbo(0.16), aids(0.16), ngos(0.16), prevention(0.15), cbos(0.14), governmental organization(0.14), program(0.13), international(0.12), intervention(0.12), organization cbos(0.12), health care(0.11), hiv aids(0.11), staff(0.11), organization(0.11), medical(0.11), delivery(0.11), improve(0.10), partnership(0.10), hiv prevention(0.10), public health(0.10), background(0.10), ngo(0.10), disease(0.10), organization ngos(0.10), implementation(0.10), conduct(0.10), method(0.09), africa(0.09), patient(0.09), nongovernmental(0.09), health service(0.09), capacity(0.08), evaluation(0.08), surgical(0.08), implement(0.08), volunteerism(0.08), country(0.08), nongovernmental organization(0.08), surgery(0.08), income country(0.07), provider(0.07), qualitative(0.07), conclusion(0.07), evidence base(0.07), health system(0.07), healthcare(0.07), training(0.07), global(0.07), organizations(0.07), community health(0.07), effective(0.06), middle income(0.06), programme(0.06), collaboration(0.06), interview(0.06), conclusions(0.06), need(0.06), barrier(0.06), epidemic(0.06), include(0.06), clinical(0.06), global health(0.06), national(0.05), base organizations(0.05), deliver(0.05), organisation(0.05), improve health(0.05), health policy(0.05), capacity building(0.05), governmental organisation(0.05), income(0.05), population(0.05), african(0.05), disparity(0.05), effectiveness(0.05),

funding(0.05), non(0.05), primary(0.05), technical(0.05), low- middle(0.04), low-(0.04), promotion(0.04), technical assistance(0.04), setting(0.04), structured(0.04), middle(0.04), experience(0.04), burden(0.04), lack(0.04), semi structured(0.04), structured interview(0.04),
**Titles of the top 10 most central articles**:
1. The Potential Of Health Sector Nongovernmental Organizations - Policy Options
2. Public-Non-Governmental Organisation Partnerships For Health: An Exploratory Study With Case Studies From Recent Ghanaian Experience
3. Non-Governmental Organizations In International Health: Past Successes, Future Challenges
4. Survey Of Nongovernmental Organizations Providing Pediatric Cardiovascular Care In Low- And Middle-Income Countries
5. Strategies To Strengthen Non-Governmental Organizations' Participation In The Iranian Health System
6. A Novel, Bottom-Up Approach To Promote Evidence-Based Hiv Prevention For People Who Inject Drugs In Ukraine: Protocol For The Mict ('bridge') Hiv Prevention Exchange Project
7. Financial Contributions To Global Surgery: An Analysis Of 160 International Charitable Organizations
8. Programmes, Resources, And Needs Of Hiv-Prevention Nongovernmental Organizations (Ngos) In Africa, Central/Eastern Europe And Central Asia, Latin America And The Caribbean
9. The Global Impact Of Surgical Volunteerism
10. Surgical Non-Governmental Organizations: Global Surgery's Unknown Nonprofit Sector

***Cluster 19***
**Title [Round 1]**: The Dynamics of Gift Giving: Cultural and Consumer Perspectives
**Title [Round 2]**: Cultural and Psychological Dynamics of Gift Givingg
**Description [Round 1]**: This cluster explores the multifaceted phenomenon of gift giving, integrating cultural, anthropological, and consumer behavior perspectives. It delves into the rituals and exchanges involved in gift processes, examining how gifting impacts social relationships and consumer motivations. By analyzing various socio-economic systems such as the gift economy and observing practices across different cultures, including those in China, the research reveals how gifts serve as both symbolic and utilitarian commodities. The cluster also investigates the role of marketing and consumer psychology in gift selection, the implications of reciprocity, and the influence of personal and cultural occasions such as Christmas on gift-giving practices. Furthermore, it considers the socio-economic implications, including the potential for gifts to function as social control or corruption tools within networks like guanxi. The research collectively aims to deepen understanding of the intricacies involved in gift exchanges and their broader socio-economic impacts.
**Description [Round 2]**: This cluster examines the multifaceted practice of gift giving through various lenses such as anthropology, consumer behavior, and social relationships. Key topics include the motivations behind giving gifts, the reciprocal nature of gift exchanges, and how gifts solidify social ties. The cluster also explores cultural variations in gift-giving practices, including specific case studies in regions like China and the Netherlands. Marketing influences and the role of consumer behavior in the gift-purchasing process are also significant subjects of discussion. Overall, this cluster provides a comprehensive insight into the complexities and underlying motivations driving the act of exchanging gifts across different cultures and contexts.
**Top 100 grams (The number in parentheses is the value of relative frequency.)**:

gift(0.83), gift giving(0.56), giving(0.53), gift give(0.28), exchange(0.26), give(0.25), consumer(0.23), giver(0.19), gift exchange(0.17), recipient(0.13), gifting(0.12), consumption(0.11), gift giver(0.10), purchase(0.09), reciprocity(0.08), gift economy(0.08), marketing(0.07), ritual(0.06), consumer behavior(0.06), receiver(0.05), economy(0.05), gift recipient(0.05), receive(0.05), giving gift(0.05), gift gift(0.05), giver recipient(0.05), occasion(0.05), friend(0.05), china(0.05), symbolic(0.05), social exchange(0.04), give behavior(0.04), buy(0.04), give gift(0.04), brand(0.04), self gift(0.04), exchange gift(0.04), type gift(0.04), chinese(0.04), consumer gift(0.03), giver receiver(0.03), sharing(0.03), receive gift(0.03), interpersonal(0.03), giving social(0.03), exchange theory(0.03), romantic(0.03), giver gift(0.03), product(0.03), user(0.03), retailer(0.03), gift purchase(0.03), choose gift(0.03), obligation(0.03), examine gift(0.03), gift behavior(0.03), study gift(0.03), mauss(0.03), customer(0.03), sale(0.03), object(0.03), recipient preference(0.03), current research(0.03), economy gift(0.03), gift social(0.03), give literature(0.03), value gift(0.03), anthropological(0.03), theory gift(0.03), meaning(0.03), commodity(0.03), holiday(0.03), feel(0.03), closeness(0.03), context gift(0.02), buyer(0.02), gift receive(0.02), relationship gift(0.02), effect gift(0.02), streaming(0.02), give practice(0.02), christmas(0.02), market exchange(0.02), utilitarian(0.02), digital(0.02), store(0.02), platform(0.02), pleasure(0.02), gift experience(0.02), gift choice(0.02), romantic gift(0.02), select gift(0.02), gift selection(0.02), retail(0.02), culture gift(0.02), purchasing(0.02), corruption(0.02), cigarette(0.02), guanxi(0.02), love(0.02),

**Titles of the top 10 most central articles**:

1. Gift Giving In Anthropological Perspective
2. Gift Giving In Hong Kong And The Continuum Of Social Ties
3. Gift Giving And The Evolution Of Cooperation
4. Gift Giving - Consumer Motivation And The Gift Purchase Process
5. Selling, Sharing, And Everything In Between: The Hybrid Economies Of Collaborative Networks
6. Thought That Counts - Signed Digraph Analysis Of Gift-Giving
7. Qualitative Steps Toward An Expanded Model Of Anxiety In Gift-Giving
8. Reciprocity As A Principle Of Exclusion: Gift Giving In The Netherlands
9. Rule Enforcement Without Visible Means - Christmas Gift Giving In Middletown
10. Gifts And Social Relations - The Mechanisms Of Reciprocity

***Cluster 20***

**Title [Round 1]**: Altruism in Organ and Gamete Donation: Motivations and Ethical Considerations

**Title [Round 2]**: Ethical Dimensions and Motivations in Altruistic Organ and Gamete Donation

**Description [Round 1]**: This cluster explores the role of altruism in the context of organ and gamete donation, examining the motivations behind such acts, as well as the associated ethical considerations. It delves into discussions about altruistic behavior, the distinction between directed and non-directed donations, and the potential role of monetary incentives in donation practices. Ethical debates surrounding the recruitment of gamete donors, the social and medical implications of organ and tissue transplantation, and the impact of different motivational appeals on donation rates are highlighted. The cluster also considers the psychosocial aspects of being a donor, including gender influences and the balance between solidarity, altruism, and effective altruism in promoting organ and gamete donations.

**Description [Round 2]**: This cluster explores the complex interplay of ethical considerations, personal motivations, and societal implications associated with altruistic donation in the contexts of organ and gamete transfers. Key discussions include the moral and ethical dynamics of altruism, especially in deciding to become a donor, whether for live organs like kidneys or reproductive materials such as sperm and eggs. The literature examines different forms of altruism, including effective altruism, and debates the potential ethical conflicts that arise when donations are motivated by prosocial appeals versus personal gain. The cluster also considers the ethical boundaries of compensation, the role of consent, and the recipient's perspective in the transplantation process. These discussions are crucial in shaping policies and ethical frameworks for organ procurement and donation programs, taking into account factors like gender, familial relationships, and societal expectations.
**Top 100 grams (The number in parentheses is the value of relative frequency.)**: donation(0.43), altruism(0.36), donor(0.36), altruistic(0.36), organ(0.29), live(0.22), kidney(0.22), donate(0.22), transplantation(0.22), transplant(0.21), recipient(0.19), organ donation(0.16), ethical(0.15), medical(0.14), kidney donation(0.13), patient(0.13), gift(0.13), living(0.12), ethic(0.11), kidney donor(0.11), live kidney(0.10), live donor(0.10), altruistic donation(0.10), exchange(0.09), consent(0.09), donor recipient(0.09), clinical(0.09), woman(0.08), kidney transplantation(0.08), shortage(0.08), reason(0.08), reproductive(0.07), accept(0.07), effective altruism(0.07), surrogacy(0.07), directed(0.07), blood(0.07), motivation(0.07), egg(0.07), health(0.06), non directed(0.06), altruistic donor(0.06), reproduction(0.06), tissue(0.06), sperm(0.06), moral(0.06), deceased(0.06), payment(0.06), organ transplantation(0.06), conclusion(0.06), ethics(0.05), oocyte(0.05), donation live(0.05), living kidney(0.05), gamete(0.05), kidney transplant(0.05), donor kidney(0.05), renal(0.05), donation altruistic(0.05), procurement(0.05), anonymous(0.05), living donor(0.05), pair(0.05), background(0.05), altruistic living(0.05), live donation(0.05), deceased donor(0.05), assist(0.05), incompatible(0.04), oocyte donation(0.04), nondirected(0.04), debate(0.04), express(0.04), motive(0.04), stranger(0.04), willing(0.04), desire(0.04), kidney exchange(0.04), donate kidney(0.04), research participation(0.04), participate(0.04), assisted(0.04), surrogate(0.04), donation altruism(0.04), recruitment(0.04), family(0.04), sperm donation(0.04), sperm donor(0.04), egg donation(0.04), body(0.04), altruistic kidney(0.04), compensation(0.04), risk(0.04), donation program(0.04), direct donation(0.04), donor donate(0.04), list(0.03), fertility(0.03), treatment(0.03), psychosocial(0.03),
**Titles of the Top 10 most central articles**:
1. Beyond The Altruistic Donor: Embedding Solidarity In Organ Procurement Policies
2. How Altruistic Organ Donation May Be (Intrinsically) Bad
3. Effective Altruists Ought To Be Allowed To Sell Their Kidneys
4. Effective Altruism And Its Critics
5. Altruism In Organ Donation: An Unnecessary Requirement?
6. Should Altruism, Solidarity, Or Reciprocity Be Used As Prosocial Appeals? Contrasting Conceptions Of Members Of The General Public And Medical Professionals Regarding Promoting Organ Donation
7. Central Role Of Altruism In The Recruitment Of Gamete Donors
8. Altruism Or Solidarity? The Motives For Organ Donation And Two Proposals
9. 'why Do You Want To Be A Donor?': Gender And The Production Of Altruism In Egg And Sperm Donation
10. Donor Insemination: The Gifting And Selling Of Semen

**D. ChatGPT-4o Prompt**

"As a neutral language model, your role is to provide a clear, objective, and logical description of various clusters based on a list of keywords and most important titles provided. These keywords are arranged in order of importance (from highest to lowest). Same for the sample titles. The information you provide will be used for academic purposes to better understand these clusters. REMEMBER THE KEYWORDS AND SAMPLE TITLES ARE IN ORDER OF IMPORTANCE. Each prompt contains keywords and sample titles. Your task is to create a title and a description for each cluster. The descriptions should be in English and must remain neutral, avoiding any endorsement or invitation towards these clusters. Your output should strictly follow this JSON structure: {title: <TITLE>, description: <DESCRIPTION>}. This structure represents a list of dictionary entry with title and description. Please note that the clusters have been identified using a clustering algorithm, meaning they organically formed in a paper citation network. The purpose of these descriptions is not to advertise the clusters, but to offer a neutral overview."